\documentclass[10pt,aps,prl,twocolumn,superscriptaddress,
showpacs,amsmath,amssymb,longbibliography,nobibnotes]{revtex4-2}
\usepackage{xcolor}
\usepackage{graphicx}
\usepackage{dcolumn}
\usepackage{bm}
\usepackage{amsmath}
\usepackage{subfigure}
\usepackage{float}
\usepackage{braket}
\usepackage{times}
\usepackage{comment}

\usepackage{amsmath}
\usepackage[bottom]{footmisc}
\usepackage{tikz}
\usetikzlibrary{positioning}

\usepackage{silence}
\usepackage[colorlinks,linkcolor=blue,citecolor=blue,urlcolor=blue]{hyperref}
\usepackage{ulem}

\begin{document}

\preprint{APS/123-QED}

\title{Multifractal Magnons Around Localized Defects}

\author{Adarsh Hullahalli}
\affiliation{Division of Physics and Applied Physics, School of Physical
and Mathematical Sciences, Nanyang Technological University,
637371 Singapore}
\affiliation{Laboratoire de Physique de l'École Normale Supérieure,
ENS, Université PSL, CNRS, Sorbonne Université, Université de Paris,
F-75005 Paris, France}

\author{Yusuf Khand}
\affiliation{Division of Physics and Applied Physics, School of Physical
and Mathematical Sciences, Nanyang Technological University,
637371 Singapore}
\affiliation{Trinity College, Cambridge, CB2 1TQ, United Kingdom}

\author{Christos Panagopoulos}
\email{christos@ntu.edu.sg}
\affiliation{Division of Physics and Applied Physics, School of Physical
and Mathematical Sciences, Nanyang Technological University,
637371 Singapore}

\author{Christina Psaroudaki}
\email{christina.psaroudaki@phys.ens.psl.eu}
\affiliation{Laboratoire de Physique de l'École Normale Supérieure,
ENS, Université PSL, CNRS, Sorbonne Université, Université de Paris,
F-75005 Paris, France}

\begin{abstract}
Multifractal wavefunctions fluctuate over a hierarchy of spatial scales, remaining spatially extended while exhibiting strongly nonuniform structure. We show that magnons in otherwise crystalline two-dimensional magnets can exhibit such behavior on mesoscopic scales, without disorder or quasiperiodicity. A localized defect can mix discrete groups of nearly degenerate magnon states in a finite sample, generating multiscale interference patterns and multifractal wavefunctions. We demonstrate this mechanism for both a simple spin defect and topological magnetic skyrmions. Large-scale spin-wave calculations for skyrmions in chiral and frustrated magnets reveal clear multifractal diagnostics over the full accessible range of system sizes of above-gap modes. Crucially, skyrmions make these states experimentally observable by hybridizing them with magnetically active internal deformation modes, which gives a subset of the multifractal states a nonzero magnetic dipole moment. These states can therefore be excited by uniform ac fields and remain distinguishable at realistic frequency resolution, providing a direct route to observing mesoscopic magnon multifractality.\end{abstract}
\maketitle

\section{\label{sec:level1} Introduction}

Extended-wave spectra in finite crystals organize into discrete energy shells whose degeneracies vary irregularly and can become arbitrarily large along sequences of increasing system size. Point-scatterer problems in singular billiards and flat tori have shown that a local perturbation can reveal the arithmetic structure of such spectra through intermediate spectral statistics, nonergodic or scarred eigenfunctions, and nontrivial small-scale equidistribution \cite{Yesha2020,PhysRevE.59.R1315,PhysRevLett.91.134103,Kurlberg2017}.

Multifractal eigenstates represent a distinct form of emergent complexity, most often associated with disorder-driven criticality \cite{PhysRevLett.67.607, aoki1983critical, PhysRevB.31.6146, schreiber1990fractal, aoki1986fractal, fal1995multifractality, vasquez2008multifractal, obuse2008boundary, huckestein1995scaling} or quasicrystalline order \cite{jagannathan2021fibonacci, mace2016fractal, holzer1991multifractal, chen2024fractal, roy2018multifractality}, and linked to nonergodicity \cite{kravtsov2015random, 10.21468/SciPostPhysCore.2.2.006, smelyanskiy2019intermittency, xu2023non, de2020multifractality}, subdiffusion \cite{10.21468/SciPostPhysCore.2.2.006, sutradhar2019transport, shimasaki2022anomalous, ohtsuki1997anomalous}, and hierarchical spatial structure. Genuine multifractality can also arise in clean short-range systems \cite{bnr3-5dcw}. Magnetic systems provide an appealing platform to search for multifractal waves. Magnons, the quanta of spin-wave excitations in ordered magnets, form the basis of magnonics and carry spin and energy without charge currents. Fractal spin-wave modes have been identified in aperiodic magnets such as quasicrystals \cite{PhysRevB.106.134431}, but whether crystalline magnets without disorder or quasiperiodicity can host accessible and controllable multifractal magnon states remains largely unknown.

Magnetic skyrmions offer a natural route to this regime. As localized topological textures in an ordered background, they break translation invariance and hybridize internal deformations with propagating magnons, generating bound states and internal resonances. In skyrmion lattices, the periodic texture acts as a magnonic crystal with bands that can acquire nontrivial topology and chiral edge modes \cite{PhysRevB.90.094423,RoldanMolina_2016,10.1126science}, while individual skyrmions produce skew scattering, magnon Hall responses, and momentum transfer \cite{PhysRevB.89.064412,PhysRevB.90.094423}. Their internal resonances are also magnetically active, enabling excitation by uniform AC fields and providing direct experimental access to hybridized states. Despite this broad phenomenology, the spatial scaling of above-gap hybridized eigenstates has remained largely unexplored. They are usually classified as extended waves or localized modes, and hybridization is analyzed in terms of resonance frequencies, selection rules, and scattering responses. 
Whether localized scattering can instead generate multifractal structure in these modes, and whether this structure can remain magnetically accessible, are open questions.

In this work, we show that a single localized defect can generate mesoscopic multifractal structure in above-gap magnon modes. The underlying mechanism relies on extended magnon states organized into near-degenerate energy shells with irregular degeneracies. By breaking translation invariance, the defect activates this spectral structure by inducing hybridization among many Fourier components across distinct length scales. Unlike disorder- or quasiperiodicity-based mechanisms, this approach requires no engineering of crystal structures or wavefunction envelopes. We establish its generality using a simple nontopological spin defect, then study skyrmions in chiral and frustrated magnets to show how dispersion and competing length scales shape the multifractal response. Finally, we show that skyrmions endow a subset of multifractal modes with nonzero net magnetic dipole moments that remain identifiable under realistic frequency resolution, and we analyze spin-current matrix elements as a complementary transport-like diagnostic of their dynamics.

\paragraph{Multifractal diagnostics.}
Multifractal structures are characterized by an infinite hierarchy of critical exponents that govern the scaling of the moments of a measure \cite{mirlin2000multifractality}. This behavior has been documented across a wide range of complex systems, including turbulent flows \cite{Boffetta_2008}, Anderson localization at metal–insulator transitions \cite {mirlin2006exact}, and nonlinear dynamical systems \cite{PhysRevA.33.1141}. To determine whether a single characteristic length scale controls a given eigenmode or if it instead displays scale-invariant spatial structure, we employ standard multifractal analysis \cite{mirlin2000multifractality}, with definitions, physical interpretation, and numerical implementation summarized in Appendix~B. For an arbitrary wavefunction $\psi_i$ on lattice sites $i$ of a $D_0$ dimensional sample, the coarse-grained observation probability in a box $n$ of side length $\ell$ is given by,
\begin{equation}\label{alpha}
P_n(\ell)=\sum_{i\in n}|\psi_i|^2 \sim \ell^{\alpha_n}\, .
\end{equation}
The singularity exponents $\alpha_n$ quantify local scaling of the measure. They are effectively uniform for monofractals and spatially varying for multifractals. Scale invariance is diagnosed by the scaling of the moments
\begin{equation}
Z(q,\ell)\equiv \sum_n P_n(\ell)^q \sim \ell^{\tau(q)} ,
\label{eq:Zqtau_main}
\end{equation}
which defines the mass exponents $\tau(q)$. For monofractals, including the fully extended limit, $\tau(q)$ is linear in $q$, whereas multifractals exhibit a characteristic nonlinearity that reflects heterogeneous scaling across space \cite{PhysRevA.33.1141}. From $\tau(q)$ we obtain the generalized dimensions $D_q=\tau(q)/(q-1)$ and the singularity spectrum $(\alpha,f(\alpha))$ through $\alpha(q)=\mathrm d\tau(q)/\mathrm dq$ and $f(\alpha)=q\alpha-\tau$ \cite{HENTSCHEL1983435}.

Throughout this work, we summarize the mode structure using two compact quantities. The information dimension $\mathcal D_1$ captures the effective dimension seen by an unbiased probe and interpolates between $\mathcal D_1=D_0$ for extended modes and $\mathcal D_1=0$ for localized modes (see Appendix C for example extended and localized modes). The spectral width $\Delta\alpha=\alpha_{\max}-\alpha_{\min}$ measures the spread of local scaling exponents that occur with nonzero weight in the spectrum and therefore quantifies heterogeneity. In the thermodynamic limit, monofractals exhibit a collapsing spectrum with $\Delta\alpha\to 0$, while genuine multifractals retain a finite width $\Delta\alpha>0$.

\begin{figure}[t]
    \includegraphics[width=1\linewidth]{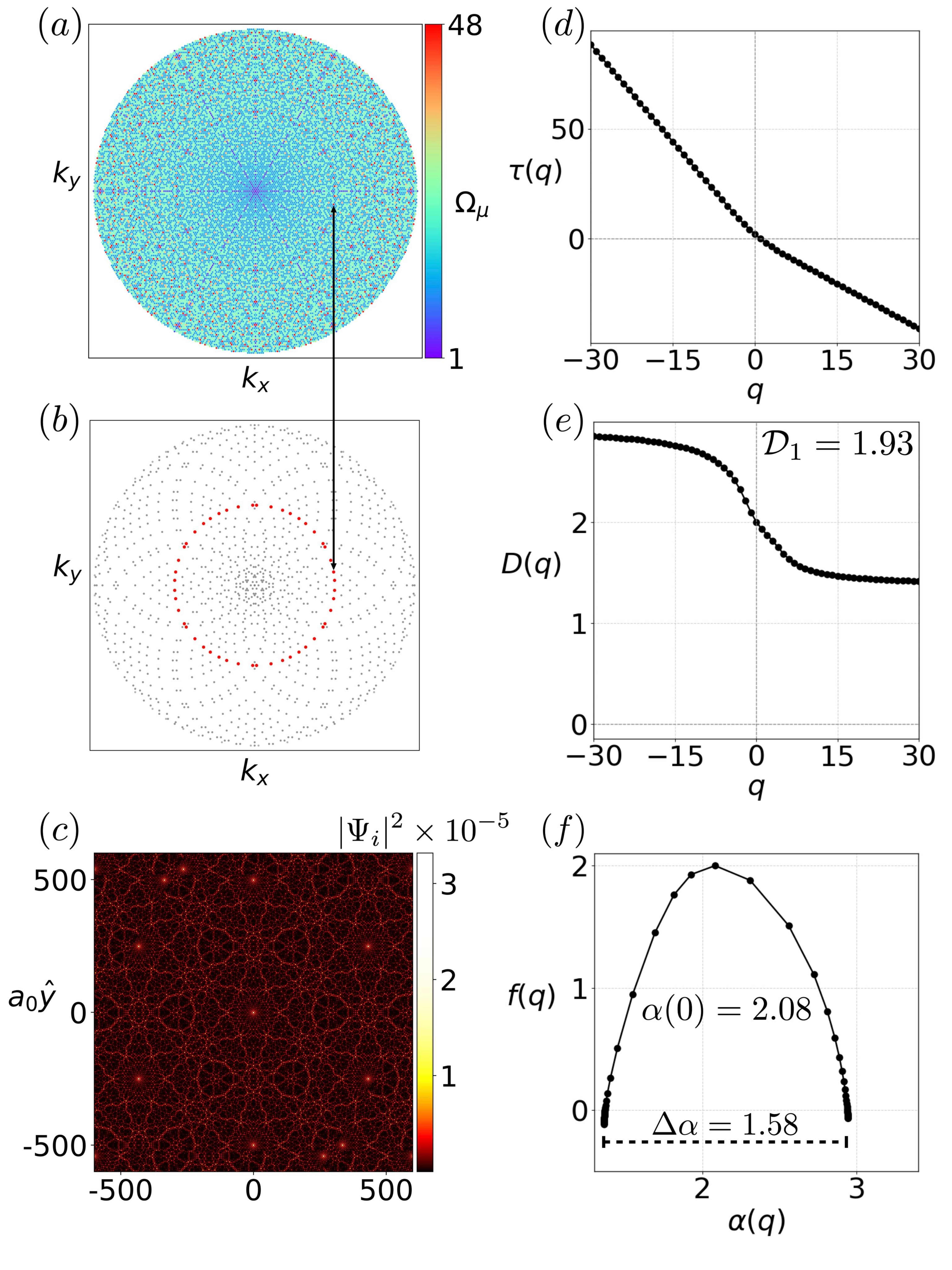}
    \caption{\textbf{Shell Structure and Multifractal Interference Patterns.} (a) First $\sim 1700$ energy level degeneracies for a chiral magnet-type dispersion. (b) Fourier components $\Delta S_\mu$ (gray dots) generated from coherent interference of states $S_\mu$ (red dots) within a chosen high-degeneracy shell ($\Omega = 48$). (c) Multifractal patterns generated by an equal-phase superposition of red states, characteristic of a defect-hybridized magnon state. (d)-(f) nonlinear $\tau(q)$, $q$-dependent generalized dimensions $\mathcal{D}_q$ with $\mathcal{D}_1<D_0$, and finite-width singularity spectrum ($\alpha$, $f(\alpha$)) confirming multifractality of the wavefunction.}
    \label{interference}
\end{figure}

\paragraph{Spectral structure and interference mechanism.}
We first show how multifractal magnon patterns can originate from the spectral organization of extended waves in a finite periodic crystal. The squared norm of any reciprocal vector on an $L\times L$ lattice with reciprocal primitive vectors $\mathbf b_i$ can be written as $|\mathbf k|^2=\sum_{i,j=1}^{D_0} n_i g_{ij} n_j$ with $ g_{ij}=\mathbf b_i\!\cdot\!\mathbf b_j$.
Introducing a scale factor $\lambda$ such that $|\mathbf k|^2=\lambda^{-1}m$ with $m\in\mathbb Z$, this becomes
\begin{equation}\label{int_norms}
m=\lambda \sum_{i,j=1}^{D_0} n_i g_{ij} n_j \, .
\end{equation}
For a dispersion $E(\mathbf k)$, eigenstates form energy shells
$S_\mu=\{\mathbf k:E(\mathbf k)=E_\mu\}$ with degeneracy $\Omega_\mu=|S_\mu|$, where $\mu$ indexes allowed values of $m$. The shell $S_{\mu}$ is shown in Fig.\ref{interference} (b) as red points. If $E(\mathbf k)=E(|\mathbf k|)$ in the energy window of interest, the shells are indexed by the discrete values of $|\mathbf k|^2$, or equivalently by the integer $m$ in Eq.~\eqref{int_norms}. Then $\Omega_\mu$ counts the number of integer solutions of the quadratic form at fixed $m$. These numbers vary irregularly with $m$, since most shells have low degeneracy, while rare high-degeneracy shells occur, and the running maximum of $\Omega_\mu$ grows without bound as the accessible range of $m$ increases, as illustrated in Fig.~\ref{interference}(a).

\begin{figure*}[t]
    \centering
    \includegraphics[width=1\linewidth] {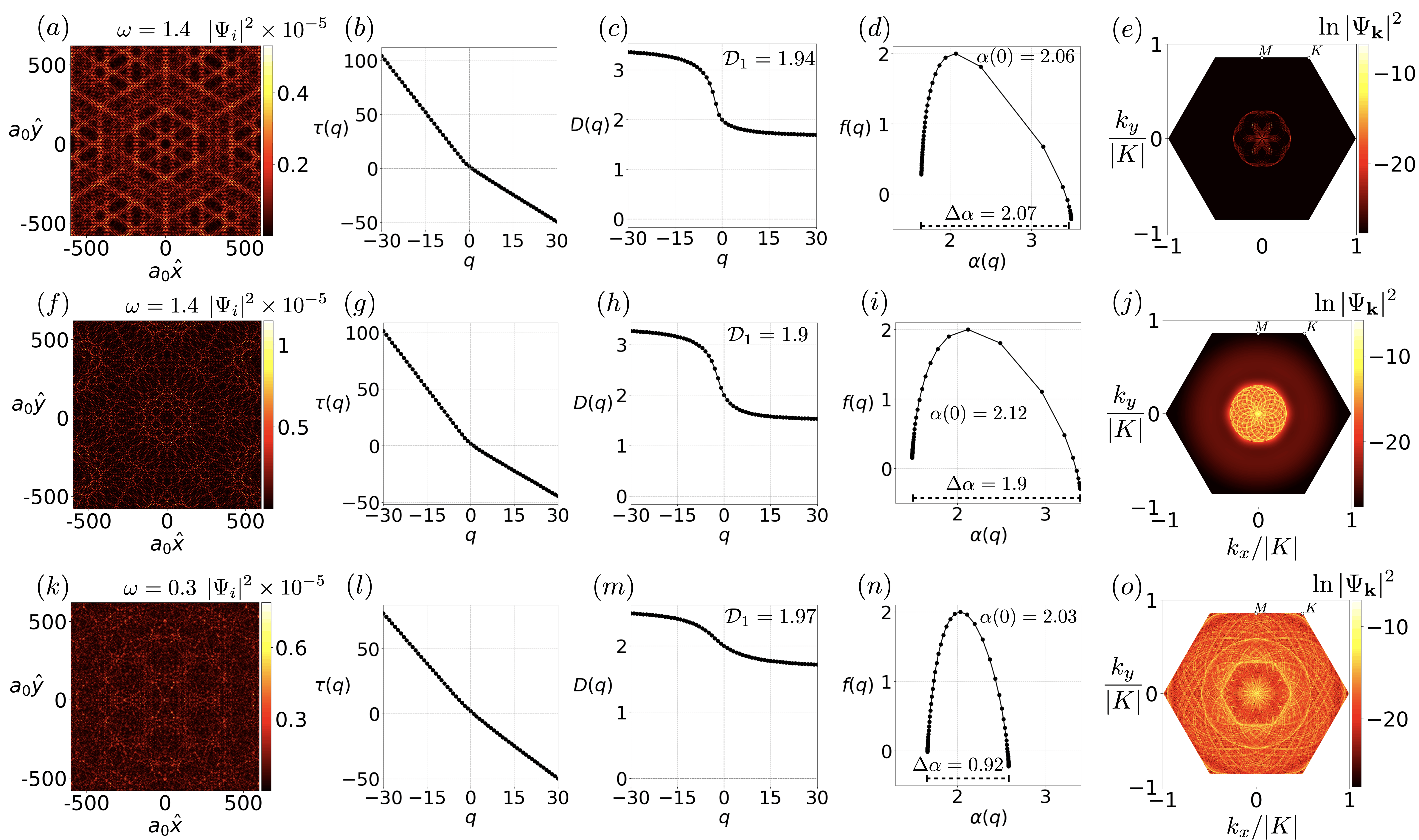}
    \caption{\textbf{Multifractal Magnons around Local Defects.} Selected multifractal modes around local defects calculated on a $1200\times 1200$ lattice, and corresponding metrics confirming multifractality. (a) Probability density of a magnon wavefunction found around an $S=0$ spin defect. (b)-(e) Respectively, nonlinear $\tau(q)$ curve, $q$-dependent generalized dimensions with $\mathcal{D}_1 < D_0$, finite-width singularity spectrum with $\alpha(0) > D_0$, and reciprocal space support. (f)-(j) Analogous plots around a chiral skyrmion. (k)-(o) Analogous plots around a frustrated skyrmion, indicating present but weaker multifractality due to a competing single-scale hybridization channel (see the high-intensity ring with radius $\sim |K|/3$ in (o)).}
    \label{skyrmion_dynamics}
\end{figure*}


A local scatterer separates the states within each shell into defect-coupled and defect-decoupled sectors, as detailed in Appendix~D. Both consist of superpositions of momenta in $S_\mu$, so their real-space intensities contain Fourier components drawn from the same difference set $\Delta S_\mu$. In the defect-decoupled sector the coefficients are basis-dependent and constrained to destructively interfere at the defect, while the defect-coupled state takes the particularly simple equal-phase form
\begin{equation}\label{coherent_wavefunction}
\psi_i=\frac{1}{\sqrt{N\Omega_\mu}} \sum_{\mathbf k\in S_\mu}
e^{i\mathbf k\cdot(\mathbf r_i-\mathbf r_0)} .
\end{equation}
Its intensity has the exact Fourier decomposition
\begin{equation}
|\psi_i|^2=
\frac{1}{N\Omega_\mu}
\sum_{\mathbf q\in\Delta S_\mu}
M(\mathbf q)e^{i\mathbf q\cdot(\mathbf r_i-\mathbf r_0)},
\end{equation}
where $\Delta S_\mu=\{\mathbf q=\mathbf k-\mathbf k' \mid \mathbf k,\mathbf k'\in S_\mu\}$ is the interference spectrum shown as grey points in Fig.~\ref{interference}(b), and $M(\mathbf q)$ counts the pairs $(\mathbf k,\mathbf k')$ producing the same $\mathbf q$. The real-space pattern is therefore controlled by the weighted interference set $\{\mathbf q,M(\mathbf q)/\Omega_\mu\}$. High-degeneracy shells generate interference components over a broad range of $|\mathbf q|$, hence many length scales, while retaining sparse reciprocal-space support because $\Delta S_\mu$ is built from a finite shell. The resulting non-self-averaging interference pattern can already be multifractal, as shown by the reconstructed wavefunction and nonlinear $\tau(q)$ in Fig.~\ref{interference}(c),(d), together with the $q$-dependent $D_q$ and finite-width singularity spectrum in Fig.~\ref{interference}(e),(f). The defect-coupled sector can additionally hybridize with nearby shells, as discussed in Appendix D.

This construction describes \textit{mesoscopic} rather than asymptotically stable multifractality. Although high shell degeneracies grow with system size, they increase more slowly than any power of $L$. Including hybridization between nearby shells, the total number $N_k$ of participating momentum states is likewise at most subpolynomial in $L$. The corresponding deviations of the positive-order dimensions from the extended value therefore vanish in the strict thermodynamic limit, with a slow crossover controlled by $\ln N_k/\ln L$. The discrete-shell interference nevertheless remains pronounced over experimentally relevant finite sizes. Its asymptotic scaling is derived in Appendix~D.

We next test this mechanism in microscopic spin-wave models. We first consider a point-like $S=0$ defect in an otherwise uniform ferromagnet, and confirm the multifractal structure expected from shell interference. We then study the effect around magnetic skyrmions, where hybridization with magnetically active internal deformations can transfer dipole weight to multifractal states. Comparing chiral and frustrated magnets reveals how the underlying dispersion controls the strength of the resulting mesoscopic multifractal response.

\paragraph{Multifractal magnons around local defects.} We consider classical spins on a $D_0=2$ dimensional triangular lattice with periodic boundary conditions. The $S=0$ defect and the chiral skyrmion are studied within the same chiral ferromagnetic model with nearest-neighbor exchange $J$, Dzyaloshinskii--Moriya interactions $D$ \cite{diaz2020chiral}, and a field along $h \parallel \hat z$ (see Eq. \ref{eq:HamChiral}). The chosen parameters $J = D = -1$, $h = -1.2$ support isolated skyrmions as long-lived metastable textures above the uniformly polarized ground state. The magnon spectrum and eigenmodes are obtained by linearizing fluctuations around the classical spin configuration, retaining quadratic terms to obtain a bosonic Bogoliubov--de Gennes-type spin-wave Hamiltonian $\mathcal H^{\rm SW}$. To access eigenstates in the vicinity of selected energies in very large systems, we use a shift-invert Arnoldi scheme \cite{arnoldi1951principle} on lattices up to $N=1200^2$ sites. 

\begin{figure}[t]
    \centering
    \includegraphics[width=1\linewidth]{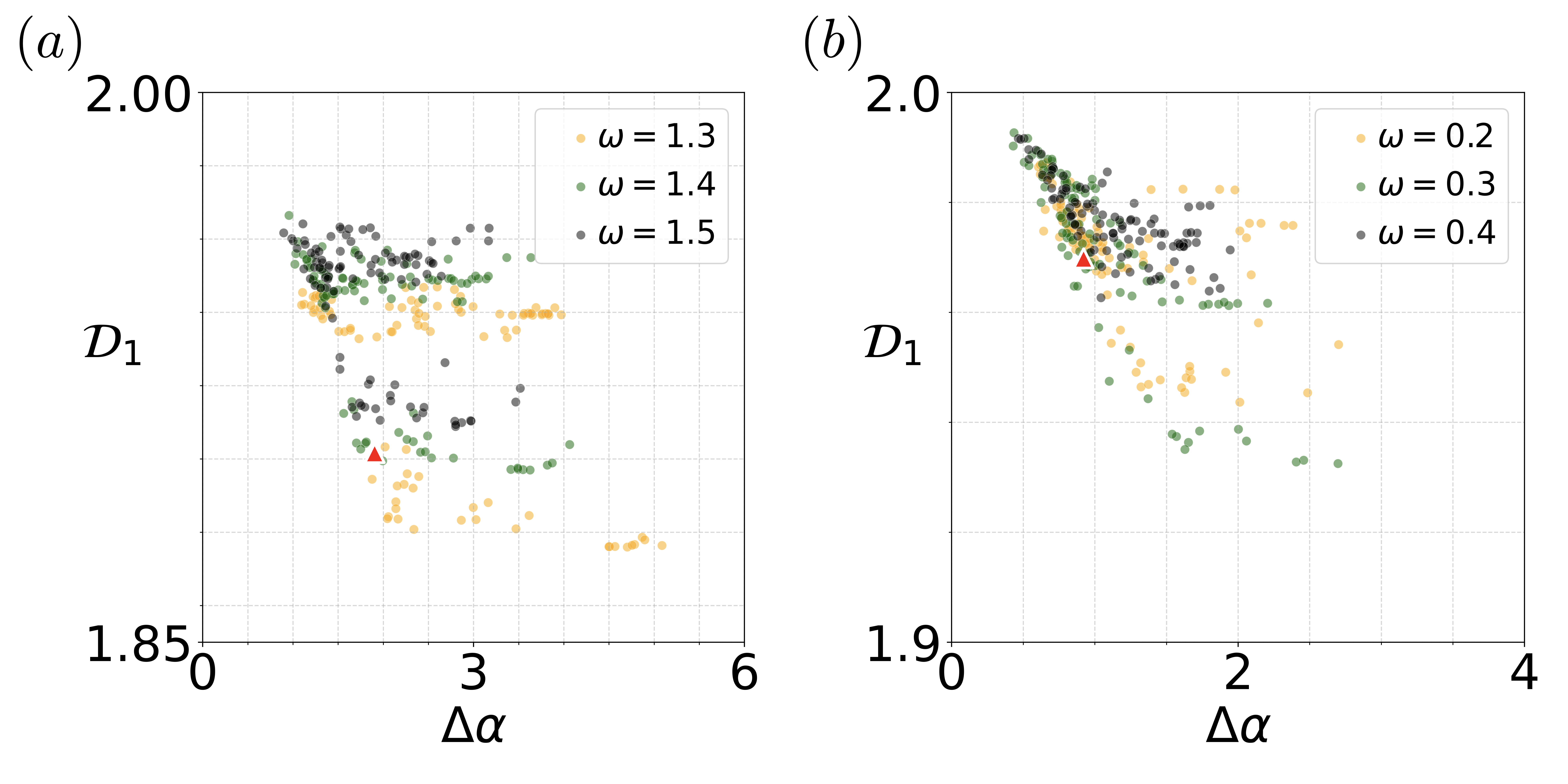}
    \caption{\textbf{Multifractal Statistics.} Scatterplots of $\Delta \alpha$ and $\mathcal{D}_1$ of the 100 modes nearest to different energy targets around a single chiral skyrmion (a) and a single frustrated skyrmion (b), illustrating how the distribution of multifractal signatures depends on the scattering regime. Frustrated modes are statistically closer to the extended limit of $\mathcal{D}_1 = 2$ and $\Delta \alpha = 0$ due to competing single-scale hybridization. Red triangles in both plots correspond to modes displayed in Fig. \ref{skyrmion_dynamics}. The chiral modes around $\omega = 1.3,\,1.4,\,1.5$ are respectively found within energy windows of width $1.47,\,0.93,\,1.41 \times 10^{-3}\,J$, and the frustrated modes around $\omega = 0.2,\,0.3,\,0.4$ within window widths of $1.20,\,3.04,\,1.08 \times 10^{-4}\,J_1$.}
    \label{Mode_Statistics}
\end{figure}

Fig. \ref{skyrmion_dynamics}(a)-(e) shows a representative magnon eigenstate obtained around a defect at $\omega = 1.4$ (with $\omega_{\rm gap} = 1.2$) and corresponding multifractal metrics. The state displays nonlinear $\tau(q)$, $\mathcal{D}_1 < D_0$, finite spectral width $\Delta \alpha$, and reciprocal-space weight sparsely distributed over multiple length scales. The defect-scattering picture suggests that multifractal structure should emerge broadly across magnon eigenstates. Consistent with this expectation, multifractal signatures are generically obtained around modes computed near several energy targets, (see Appendix~F).

We next replace the spin defect with a chiral skyrmion, and similarly observe multifractality in the above-gap skyrmion-scattered eigenstates as illustrated in Fig. \ref{skyrmion_dynamics}(f)-(j)). In contrast, below-gap localized modes of the skyrmion remain localized, while magnons in the uniform ferromagnet recover the extended limit. The multifractal response is not restricted to lattice-scale skyrmions and persists as the skyrmion size is increased. It also survives in the presence of multiple skyrmions as the scattering from additional skyrmions scrambles the simple coherent interference pattern of an isolated defect, but does not remove the underlying multifractal structure. Further numerical checks of skyrmion-size dependence and multiple-skyrmion configurations are given in Appendices~E and G.

We additionally consider a frustrated magnet with competing $J_1$--$J_2$ exchange, easy-axis anisotropy $\kappa$ and field $h \parallel \hat z$ \cite{Leonov2015}, with parameters $J_1 = -1$, $J_2 = 0.5$, $h = -0.3$, $\kappa = -0.15$ supporting metastable isolated skyrmions and setting $\omega_{\rm gap} = 0.1$. While the $S=0$ defect and chiral skyrmion are well-approximated as point scatterers, a qualitatively different regime emerges in the frustrated platform, where the Mexican-hat dispersion introduces a preferred magnon wavelength set by the frustration length and comparable to the skyrmion size. This opens an efficient hybridization channel with finite-wavelength magnons and produces a pronounced single-scale structure \cite{fhbj-ykvn} while leaving weaker residual multifractal fluctuations. The representative mode obtained at $\omega = 0.3$ shown in Fig. \ref{skyrmion_dynamics}(k)-(o) reflects this competition, exhibiting multifractal diagnostics shifted toward the extended limit and a high-intensity reciprocal-space ring at $|\mathbf{k}|\sim |K|/3$, corresponding with the skyrmion size.

To systematically characterize the influence of the magnon dispersion, we analyze $100$ eigenstates around energy targets $\omega = \omega_{\rm gap} + 0.1, 0.2, 0.3$ in both skyrmion systems. As shown in Fig. \ref{Mode_Statistics}, multifractal signatures $\mathcal{D}_1 < D_0$ and finite $\Delta \alpha$ persist across all calculated modes. However, the frustrated modes are statistically closer to the extended limit $\mathcal{D}_1 = D_0$ and $\Delta \alpha = 0$, confirming that the competing single-scale channel weakens but does not eliminate multifractal signals.

\begin{figure}[t]
\centering
\includegraphics[width=1\linewidth]{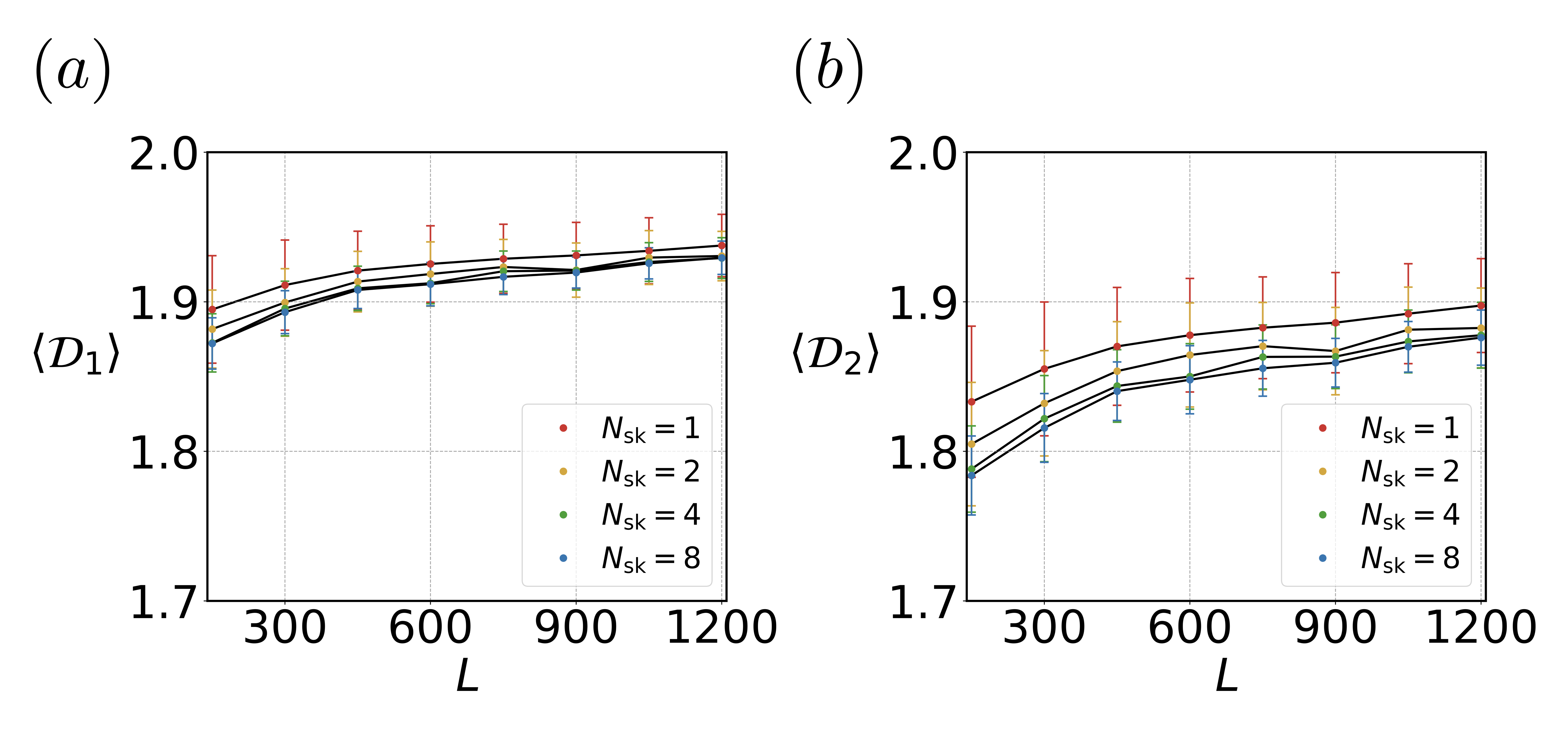}
\caption{\textbf{System Size and Skyrmion Number Dependence.} (a) Information dimension $D_1$ and (b) correlation dimension $D_2$, averaged over 100 chiral magnon modes, versus linear system size $L$. Each curve corresponds to a fixed number $N_{\mathrm{sk}}$ of randomly positioned skyrmions. Both dimensions remain below the extended limit $D_1=D_2=2$ over mesoscopic sizes.}
    \label{fig:D1_D2_L}
\end{figure}

Finally, Fig.~\ref{fig:D1_D2_L}(a),(b) shows that, for fixed skyrmion number, $\mathcal{D}_1$ and $\mathcal{D}_2$ approach $D_0$ logarithmically with system size, while remaining below $D_0$ up to $L=1200$. The states thus retain multifractal structure at realistic mesoscopic sizes. At fixed $L$, adding skyrmions enhances multifractality, but the effect rapidly saturates, with little change between $N_{\mathrm{sk}}=4$ and $8$. A complementary phase-randomization test
[Fig.~\ref{Phase_Randomization} of the Appendix] shows that increasing phase randomness raises both dimensions toward $D_0$, before they saturate below the extended limit. Phase scrambling weakens, but does not eliminate, mesoscopic multifractality. This trend supports competition between stronger shell hybridization and increasing phase scrambling by additional scatterers.

\paragraph{Observability.} We now consider the experimentally accessible subset of modes hybridized with magnetically active internal deformations of the skyrmion. This hybridization transfers dipole weight $\mathcal M_n^\nu$ to above-gap resonances, allowing them to couple to uniform AC fields \cite{PhysRevLett.108.017601,PhysRevLett.109.037603} (see Fig. \ref{Observable_Modes} in Appendix~A). Representative accessible wavefunctions selected by large dipole moment are included in Appendix~J, confirming that accessible modes retain characteristic multifractal spatial structure and typical metrics. 

\begin{figure}[t]
\includegraphics[width=0.9\linewidth]{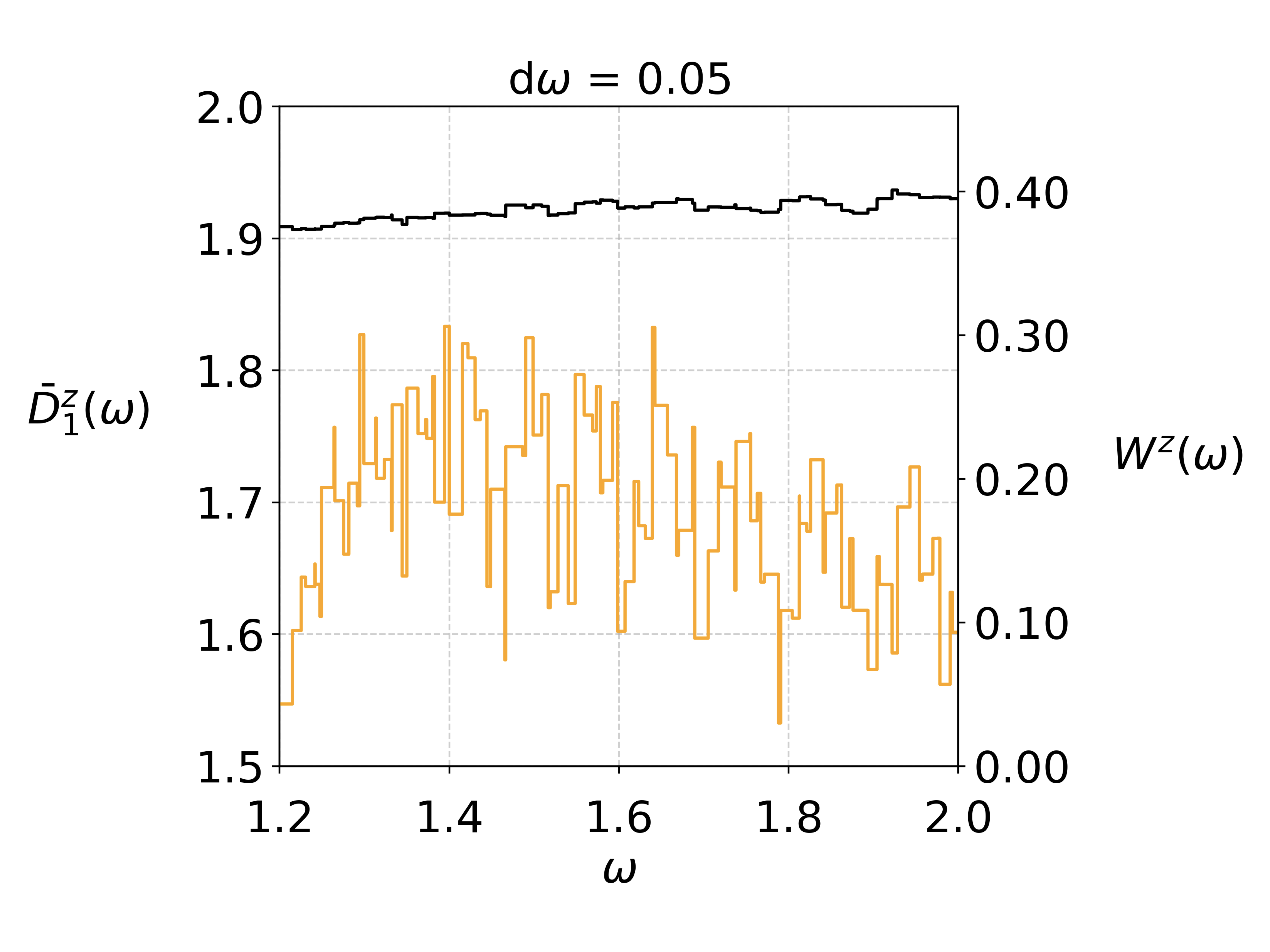}
\caption{\textbf{Observable Multifractal Magnons.}
Longitudinal dipole spectral weight $W^z(\omega)$ and dipole-weighted information dimension $\overline{D}^{\,z}_1(\omega)$ for the full magnon spectrum of a $120\times120$ chiral system, with frequency resolution
$\delta\omega=0.05$. Windows with finite $W^z$ and $\overline{D}^{\,z}_1<2$ identify spectral windows populated by magnetically accessible multifractal modes.}
\label{FractDimens}
\end{figure}

In large systems, nearby dipole-active modes become closely spaced in energy and cannot be individually resolved. At a frequency resolution $\delta \omega$, we quantify the response of an ensemble of multifractal magnons by the binned spectral weight
\begin{equation}
W^\nu(\omega)=\sum_n |\mathcal M_n^\nu|^2\,\Theta\!\left(\delta\omega/2-|\omega-\omega_n|\right),
\end{equation}
and the dipole-weighted information dimension
\begin{equation}
\overline{D}^{\,\nu}_1(\omega)=
\frac{\sum_n |\mathcal M^\nu_n|^2\,\Theta\!\left(\frac{\delta\omega}{2}-|\omega-\omega_n|\right)\,D_{1,n}}
{\sum_n |\mathcal M^\nu_n|^2\,\Theta\!\left(\frac{\delta\omega}{2}-|\omega-\omega_n|\right)}.
\label{eq:D1bar}
\end{equation}
For $\delta\omega=0.05$, Fig.~\ref{FractDimens} shows that frequency windows with finite $W^z(\omega)$ still satisfy $\overline{D}^{\,z}_1(\omega)<D_0=2$. Different bandwidths $\delta \omega$ are considered in the Appendix~H, indicating that spectroscopic observability remains robust to finite frequency resolution. Complementary spin-current diagnostics are presented in Appendix~I. 

\paragraph{Experimental probes and materials.}
For realistic units $J=1$ meV, $D=0.15$ meV, and $B=hJ/(g\mu_B)=0.43$ T, the onset of multifractal excitations occurs above a gap $\omega_{\rm gap}=0.05$ meV $\simeq 12$ GHz. Our results describe effectively two-dimensional skyrmion-hosting geometries in which out-of-plane magnon modes are frozen, and are compatible with insulating platforms such as, but not limited to, $\mathrm{Cu_2OSeO_3}$ \cite{seki2012observation} and $\mathrm{GaV_4Se_8}$ \cite{fujima2017thermodynamically,tokura2020magnetic}, provided the relevant scales satisfy $R_{\rm sk}\ll \ell_{\rm sample}\lesssim \ell_{\rm c}$, with $R_{\rm sk}$ the skyrmion size, $\ell_{\rm sample}$ the sample size, and $\ell_{\rm c}$ the magnon coherence length.

Chiral magnets are especially promising for observing strong multifractal signals. Because individual mode spacing decreases with system size, the relevant observable is a frequency-binned response of nearby dipole-active modes, and the windows used in Figs.~\ref{FractDimens} and \ref{Observability_Appendix} correspond to experimental linewidths of order $1$--$10$ GHz. The resulting multiscale wavefunctions can be probed by techniques with spatial sensitivity to spin-wave structure, including color-center magnetometry \cite{bertelli2020magnetic, manas2025isofrequency}, scanning transmission X-ray microscopy \cite{sluka2019emission}, and Brillouin light scattering \cite{sebastian2015micro}. They should also be accessible to scanning probes. For spin fluctuations $\delta S_x/S\sim10^{-3}$ and $M_s=10^6$ A/m, one expects surface transverse fields $\Delta B_\perp\sim1$ mT and stray fields of order $100~\mu$T at a lift height of $10$ nm.

Beyond conventional multifractal metrics, the present states are expected to exhibit experimentally accessible signatures. In particular, multifractal wavefunctions are associated with broad, non-Gaussian distributions of local intensity, which in the present context would manifest as strong spatial fluctuations of the dynamical magnetization around a skyrmion, potentially accessible via spatially resolved probes such as NV magnetometry or scanning transmission X-ray microscopy \cite{RevModPhys.80.1355}. Furthermore, the sparse and multiscale momentum-space structure of these modes may give rise to a correspondingly inhomogeneous distribution of spectral weight in the dynamical structure factor S(q,$\omega$), providing a possible signature in Brillouin light scattering or neutron scattering measurements.

\paragraph{Conclusions.}
We have shown that localized defects in magnetic crystals can produce multifractal structure in above-gap magnon modes over mesoscopic system sizes. The underlying mechanism is defect-induced mixing of discrete groups of nearly-degenerate magnon states with irregular degeneracies, which generates multiscale interference patterns with sparse structure. A spin defect provides a simple realization of this effect, while magnetic skyrmions make it experimentally accessible through hybridization with their internal deformations. Comparing chiral and frustrated skyrmions further shows how the scattering regime shapes the multifractal response, with the competing single-scale channel in frustrated magnets statistically shifting multifractal signatures toward the extended limit. Magnetically active multifractal modes around skyrmions remain identifiable under realistic frequency resolution, while spin-current matrix elements provide a complementary transport-like signature.

More broadly, our results uncover a direct link between the spectral arithmetic of finite crystals and emergent multiscale wave structure activated by scattering from local defects. This establishes a route to multifractality in otherwise crystalline magnets without disorder or quasiperiodicity, with skyrmion systems providing a tunable and experimentally accessible realization.

\paragraph{Acknowledgments}
C.Ps. is an \'{E}cole Normale Sup\'{e}rieure (ENS)-Mitsubishi Heavy Industries (MHI) Chair of Quantum Information supported by MHI. The work in Singapore was supported by the Singapore Ministry of Education (MOE) Academic Research Fund Tier 3 grant MOE-MOET32023-0003.
\bibliographystyle{apsrev4-2}
\bibliography{references}
\newpage 

\begin{widetext}

\section*{Appendix}\label{sec:Appendix}
\subsection*{A. Magnon Dynamics}\label{App:Magnons}

We consider magnetic spins on skyrmion-hosting chiral and frustrated magnetic triangular lattices with periodic boundary conditions and energetics respectively described by the following classical Hamiltonians:

\begin{equation}
    \mathcal{H}_\text{chiral} = J \sum_{\langle ij \rangle} \mathbf{S}_i \cdot \mathbf{S}_j + D\sum_{\langle ij \rangle} \mathbf{d}_{ij} \cdot (\mathbf{S}_i \times \mathbf{S}_j) + h \sum_i \mathbf{S}_i \cdot \mathbf{\hat{z}}\label{eq:HamChiral}
\end{equation}

\begin{equation}
    \mathcal{H}_\text{frustrated} = J_1 \sum_{\langle ij \rangle} \mathbf{S}_i \cdot \mathbf{S}_j + J_2 \sum_{\langle \langle ij \rangle \rangle} \mathbf{S}_i \cdot \mathbf{S}_j + h \sum_i \mathbf{S}_i \cdot \mathbf{\hat{z}} + \kappa\sum_i (\mathbf{S}_i \cdot \mathbf{\hat{z}})^2\,,\label{eq:HamFrustrated}
\end{equation}

\noindent
where $\langle\rangle$ and  $\langle\langle\rangle\rangle$ denote nearest and next nearest neighbors. The chosen parameters are $J = -1$, $D = -1$, and $h = -1.2$ with Bloch-type DMI vectors for the chiral magnet and $J_1 = -1$, $J_2 = 0.5$, $h=-0.3$, and $\kappa=-0.15$ for a frustrated magnet, and always correspond to uniformly magnetized ground states in which metastable skyrmions can exist. We first initialize skyrmions in this uniform state and then numerically minimize the energy by solving the Landau-Lifshitz-Gilbert (LLG) equation with strong damping. Fluctuations around this stable classical configuration are calculated to linear order from the following spin-wave Hamiltonians obtained with the Holstein-Primakoff transformation in a rotating orthonormal basis $\{\bm v_i^{1},\bm v_i^{2},\bm v_i^{3}\}$ where $\bm v_i^{3}\parallel \bm S_{i,0}$, and with $\bm v_i^{+} = \bm v_i^{1} + i\bm v_i^{2}$ and $\bm v_i^{-} = \bm v_i^{1} - i\bm v_i^{2}$:

\begin{equation}\label{eq:magnon_H}
    \mathcal{H}^{\text{SW}} = \begin{pmatrix} \Phi^{-} + \mathcal{C} & \Phi^{+} \\ (\Phi^{+})^* & (\Phi^{-})^* + \mathcal{C} \end{pmatrix}\,, 
\end{equation}

\noindent
For a chiral magnet:

\begin{equation}
    \mathcal{C}_{\mu \nu} = -\delta_{\mu \nu} \left[J \sum_{\langle ij \rangle} \mathbf{v}^3_i \cdot \mathbf{v}^3_j + D\sum_{\langle ij \rangle} \mathbf{d}_{ij} \cdot (\mathbf{v}^3_i \times \mathbf{v}^3_j)+ \frac{1}{S}h \mathbf{\hat{z}} \cdot \mathbf{v}^3_i\right]_{i=\mu}
\end{equation}

\begin{equation}
    \Phi^{\pm}_{\mu \nu} = \frac{1}{2} \left[J \mathbf{v}^+_i \cdot \mathbf{v}^\pm_j + D\sum_{\langle ij \rangle} \mathbf{d}_{ij} \cdot (\mathbf{v}^+_i \times \mathbf{v}^\pm_j)\right]_{i=\mu, j=\nu}\,, 
\end{equation}

\noindent
and for a frustrated magnet:

\begin{equation}
    \mathcal{C}_{\mu \nu} = -\delta_{\mu \nu} \left[J_1 \sum_{\langle ij \rangle} \mathbf{v}^3_i \cdot \mathbf{v}^3_j + J_2 \sum_{\langle \langle ij \rangle \rangle} \mathbf{v}^3_i \cdot \mathbf{v}^3_j + 2\kappa (\mathbf{\hat{z}} \cdot \mathbf{v}^3_i)^2 + \frac{1}{S}h \mathbf{\hat{z}} \cdot \mathbf{v}^3_i\right]_{i=\mu}
\end{equation}

\begin{equation}
    \Phi^{\pm}_{\mu \nu} = \frac{1}{2} \left[J_1 \mathbf{v}^+_i \cdot \mathbf{v}^\pm_j + J_2 \mathbf{v}^+_i \cdot \mathbf{v}^\pm_j + \delta_{ij} 2\kappa (\mathbf{\hat{z}} \cdot \mathbf{v}^+_i) (\mathbf{\hat{z}} \cdot \mathbf{v}^\pm_j)\right]_{i=\mu, j=\nu}
\end{equation}

\noindent
The reported multifractal eigenstates are calculated via shift-invert Arnoldi algorithm \cite{arnoldi1951principle} from the dynamical matrix $\sigma_z \mathcal{H}^{\text{SW}}$ obtained around a single skyrmion in a crystal of $N=1200^2$ lattice sites with periodic boundary conditions nearby various energy targets. Magnon wavefunctions $\ket{\psi}$ are superpositions of particle and hole components, the latter of which occur wherever the local frame rotates - here at the skyrmion. Single-magnon orbitals are normalized as $\sum_i(\psi_{i\text{, particle}}^2 - \psi_{i\text{, hole}}^2) = 1$, but the natural definition of magnon probabilities that follows, $|\psi_i|^2 = \psi_{i\text{, particle}}^2 - \psi_{i\text{, hole}}^2$, is unsuitable for fractal analysis as it can be locally negative. We use the alternative definition:

\begin{equation}
    |\psi_i|^2 = \left(\psi_{i\text{, particle}}^2 + \psi_{i\text{, hole}}^2\right)/\sum_j \left(\psi_{j\text{, particle}}^2 + \psi_{j\text{, hole}}^2\right)\,,
\end{equation}

\noindent
which is equivalent to the former in the limit $N_{\text{uniform}} \gg N_{\text{skyrmion}}$, where $N_{\text{uniform}}$ is the size of the subset of spins without the skyrmion and $N_{\text{skyrmion}}$ is its complement. This is always the case in the reported results. 

\begin{figure}
    \centering
    \includegraphics[width=0.95\linewidth]{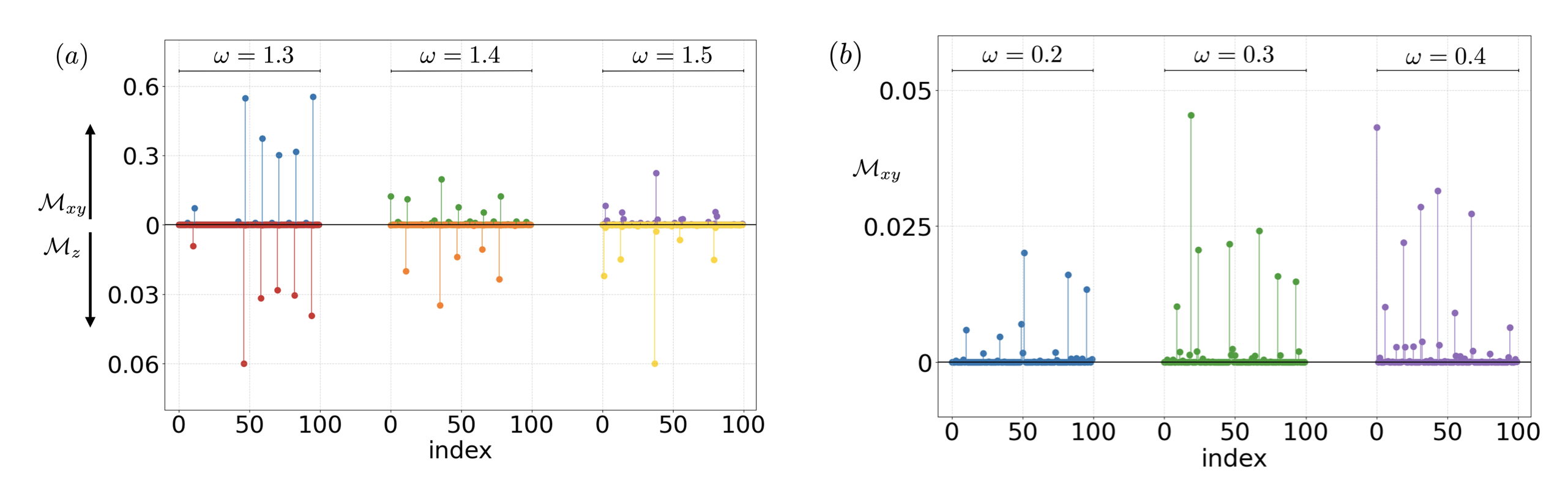}
    \caption{\textbf{Observable Modes.} Magnetic dipole moments calculated from Eq. \eqref{dipole} for magnons around (a) a single chiral skyrmion and (b) a single frustrated skyrmion. Chiral skyrmions possess both in-plane and out-of-plane moments, while centrosymmetric frustrated magnets possess only in-plane components \cite{fhbj-ykvn}. The moments quantify mode coupling to AC fields, therefore, a subset of multifractal modes are observable in both cases. Note that the magnetically active modes with large in-plane and out-of-plane moments in (a) are not the same mode, but neighboring ones.}
    \label{Observable_Modes}
\end{figure}

Components $\mathcal{M}^\nu$ of magnetic dipole moments are calculated from the following \cite{hirosawa2022magnetoelectric}:

\begin{equation}\label{dipole}
    \mathcal{M}^{\nu} = \sum_i^N \sqrt{\frac{S}{2}} \left(\mathbf{v}_i^- \psi_{i\text{, particle}} + \mathbf{v}_i^+ \psi_{i\text{, hole}}\right)
\end{equation}
Chiral skyrmions support modes with both in-plane and out-of-plane dipole moments. In contrast, for the centrosymmetric frustrated model the out-of-plane response is forbidden to linear order, and only in-plane dipole moments occur \cite{fhbj-ykvn}. Figure~\ref{Observable_Modes} summarizes the resulting dipole distributions, showing that in both models a non-negligible subset of the multifractal above-gap modes is magnetically active and can be excited by uniform AC fields.  

\subsection*{B. Multifractal Analysis}\label{App:MultifractalAnalyisis}

The coarse-grained observation probability of a wavefunction within a box indexed by $n$ of side length $\ell$ can be written as

\begin{equation}\label{alpha}
    P_n(\ell) = \sum_{i \in n} |\psi_i|^2 \sim \ell^{\alpha_n}\,,
\end{equation}

\noindent
The local singularity exponent $\alpha_n$ characterizes how the probability contained within a spatial region scales with $\ell$. For monofractal patterns, $\alpha_n$ is effectively uniform across space, whereas a multifractal contains subsets with different singularity exponents. The latter can be identified and characterized by the singularity spectrum $(\alpha, f(\alpha))$, which assigns a fractal dimension $f(\alpha)$ describing the geometric scaling of the number of boxes $N(\alpha)$ to each $\alpha$:

\begin{equation}\label{f_alpha}
    N(\alpha) \sim \ell^{-f(\alpha)}
\end{equation}
We calculate spectra from mass exponents $\tau(q)$ related to moments of the coarse-grained probability density $P^q_n(\ell),\,q \in (-\infty, \infty)$ as:

\begin{equation}
    \sum_{n=1}^N P^q_n(\ell) \sim \ell^{\tau(q)}
\end{equation}

\noindent
Which is in principle related simply to $(\alpha, f(\alpha))$ as  $\alpha(q) = d\tau(q)/dq$ and $f(\alpha(q)) = q\alpha(q) - \tau(q)$. However, for reasons of numerical accuracy, we instead calculate the spectrum from the following limits \cite{PhysRevLett.62.1327}:

\begin{equation}\label{alpha_numeric}
    \alpha(q) = \lim_{\ell\rightarrow 0} \frac{1}{\ln{\ell}} \sum_{n=1}^N \frac{P_n^q(\ell)} {\sum_{m=1}^N P_m^q(\ell)} \ln P_n(\ell).
\end{equation}

\begin{equation}\label{f_numeric}
    f(q) = \lim_{\ell\rightarrow 0} \frac{1}{\ln{\ell}}\sum_{n=1}^N \frac{P_n^q(\ell)}{\sum_{m=1}^N P_m^q(\ell)} \left(\ln{\frac{P_n^q(\ell)}{\sum_{m=1}^N P_m^q(\ell)}} \right)
\end{equation} 

\noindent
Multifractal strength is well-described by the width $\Delta \alpha$, which determines the range of distinct $\alpha$ present across subsets of the wavefunction.

From the mass exponents, we also calculate the generalized dimensions $\mathcal{D}_q$, fractal dimensions of the wavefunction's real space support as observed by a probe weighting local probabilities as $P^q_n$, obtained via

\begin{equation}\label{Dq}
    \mathcal{D}_q = \frac{\tau(q)}{q-1} =\frac{1}{q-1}\lim_{\ell \rightarrow 0} \frac{\ln{\sum_{n=1}^NP^q_n}}{\ln{\ell}}.
\end{equation}

\noindent
Positive $q$ emphasizes high-intensity regions, while negative $q$ emphasizes low-intensity regions and voids. For strongly inhomogeneous measures, these limits probe different geometric subsets, yielding dimensions well below $D_0$ on concentrated support and much larger in the void sector. The localization of a wavefunction is well-described by the information dimension $\mathcal{D}_1$, the dimension observed by an unbiased ($q=1$) probe. In the thermodynamic limit, nonfractal localized and extended modes respectively correspond to $\mathcal{D}_1 \rightarrow 0$ and $\mathcal{D}_1 \rightarrow \mathcal{D}_0$, while fractal modes take on intermediate values. 

\subsection{C. Extended and Localized Modes}\label{app:nonfractals}

For reference, we show the multifractal diagnostics for representative nonfractal extended and localized states, obtained respectively from unperturbed magnon modes and below-gap magnon–skyrmion bound states.

\begin{figure}[H]
    \centering
    \includegraphics[width=1\linewidth]{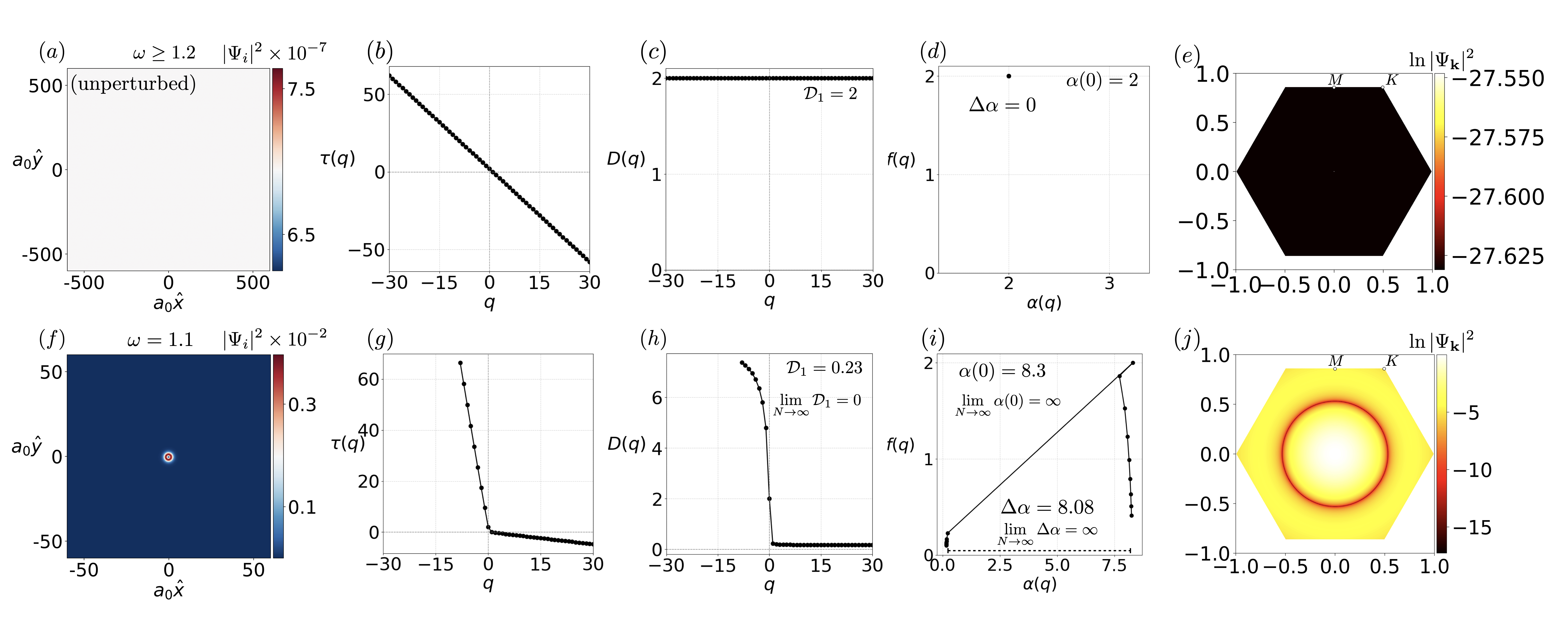}
    \caption{\textbf{Extended and Localized Modes.} Metrics reported in Fig. \ref{skyrmion_dynamics} are calculated from above-gap extended states of a chiral ferromagnet and below-gap localized states of a single chiral skyrmion. (a) Uniform probability density of magnon eigenstates around the ferromagnetic ground state. (b) Linear $\tau(q)$ curve. (c) Generalized dimensions $\mathcal{D}_q = D_0=2$, illustrating trivial scaling of all moments of the probability density. (d) Single-point singularity spectrum $(\alpha, f(\alpha))$ at $\alpha = f(\alpha) = 2$, indicating uniform and nonfractal wavefunction scaling. (e) Reciprocal space plot with only a single peak at $\mathbf{k}=0$. (f) Localized mode of a single chiral skyrmion. (g) Highly nonlinear mass exponents for $q \rightarrow -\infty$. (h) Large generalized dimensions for $q < 0$ moments corresponding to voids and information dimension $\mathcal{D}_1$ approaching zero, both indicative of localization. (i) Singularity spectrum illustrating divergence of values of $\alpha$ with nonzero fractal dimensions $f(\alpha)$ expected from a wavefunction concentrated at a single point in the thermodynamic limit. (j) Fully extended reciprocal lattice support, with the exception of a single ring corresponding to wavelengths $\sim (1/2)R_{\text{sk}}$, for which the signals cancel.}
    \label{Localized_Extended}
\end{figure}

\subsection{D. Multifractality in the thermodynamic and continuum limits}
\label{app:limits}

The thermodynamic and continuum limits correspond to distinct limiting procedures and should be considered separately. Let $a_0$ denote the lattice constant, $R_d$ the physical defect size, and $L$ the number of lattice sites along each spatial direction, so that the physical linear system size is $La_0$. The thermodynamic limit, $L\rightarrow\infty$ at fixed $a_0$ and $R_d$, preserves the discrete crystalline structure. By contrast, the continuum limit $a_0\rightarrow0$ at fixed $R_d$ and physical system size $La_0$ treats the lattice as unresolved and describes the spin fields as continuous functions. The thermodynamic analysis shows that multifractality remains possible at mesoscopic system sizes.

\paragraph{Thermodynamic limit.} Extended magnon eigenstates $\ket{\mathbf{k}}$ in two dimensions are normalized over $L^2$ lattice sites and therefore have local amplitudes $\left|\langle \mathbf{r}|\mathbf{k}\rangle\right| = L^{-1}$. We consider a quadratic dispersion, $\omega - \omega_{\rm gap} \propto |\mathbf{k}|^2$, and defects microscopic relative to low-energy magnon wavelengths. Such defects cannot resolve momentum differences between coupled magnon states within a shell and may therefore be approximated as point-like. Placing the defect at the origin, the defect operator takes the form

\begin{equation}
    \hat{V} = U\ket{0}\bra{0}
\end{equation}
for scalar defect strength $U$. The matrix elements $\hat{V}_{\mathbf{k}, \mathbf{k}'}$ are uniform and given by

\begin{equation}
    \hat{V}_{\mathbf{k}, \mathbf{k}'} = \frac{U}{L^2}.
\end{equation}

In a finite system, magnon states are organized into discrete energy shells $S_\mu$ with finite degeneracy $\Omega_\mu$. For periodic boundary conditions, each shell is associated with an integer norm

\begin{equation}
    m_\mu = Q(n_1, n_2), \qquad n_1, n_2 \in \mathbb{Z}
\end{equation}
representable by a quadratic form. For triangular and square lattices, $Q(n_1, n_2)$ are respectively given by:

\begin{equation}
    Q_{\triangle}(n_1,n_2)=n_1^2+n_1n_2+n_2^2,
    \qquad
    Q_{\square}(n_1,n_2)=n_1^2+n_2^2.
\end{equation}
The magnon dispersion may now be rewritten in terms of the arithmetic norms $m$,

\begin{equation}\label{dispersion_arithmetic}
    \omega_\mu = \omega_{\rm gap} + \frac{\lambda m_\mu}{L^2}.
\end{equation}
For shells near a fixed frequency above the gap, this implies that $m_\mu \sim L^2$. The defect operator acting on the shell states is

\begin{equation}
    V_\mu
    =
    \frac{U}{L^2}
    \begin{pmatrix}
    1 & 1 & \cdots & 1\\
    1 & 1 & \cdots & 1\\
    \vdots & \vdots & \ddots & \vdots\\
    1 & 1 & \cdots & 1
    \end{pmatrix}_{\Omega_\mu\times\Omega_\mu}.
\end{equation}

Diagonalizing $V_\mu$ separates the shell into defect-coupled and defect-decoupled sectors. The former consists of a single equal-phase superposition state which constructively interferes at the defect position,

\begin{equation}\label{coupled_single_defect}
    \left|B_\mu\right\rangle
    =
    \frac{1}{\sqrt{\Omega_\mu}}
    \sum_{\mathbf{k}\in S_\mu}
    |\mathbf{k}\rangle,
\end{equation}
and acquires an energy correction,

\begin{equation}\label{numerator_scaling}
    V_\mu |B_\mu\rangle
    =
    \frac{U\Omega_\mu}{L^2}
    |B_\mu\rangle.
\end{equation}
The latter comprises a basis of $\Omega_\mu - 1$ states which destructively interfere at the defect,

\begin{equation}
    |D_{\mu,j}\rangle
    =
    \sum_{\mathbf{k}\in S_\mu}
    c_{\mathbf{k}}^{(j)}
    |\mathbf{k}\rangle,
    \qquad
    \sum_{\mathbf{k}\in S_\mu}
    c_{\mathbf{k}}^{(j)}=0,
    \qquad
    \sum_{\mathbf{k}\in S_\mu}
    \left|c_{\mathbf{k}}^{(j)}\right|^2=1,
\end{equation}
and do not acquire any energy corrections:

\begin{equation}
    V_\mu |D_{\mu,j}\rangle = 0,
    \qquad
    j=1,\ldots,\Omega_\mu-1.
\end{equation}

The defect-decoupled sector is already diagonal in the defect Hamiltonian, and therefore does not mix with the rest of the spectrum. The number of momenta carried by a state in this sector is upper bounded by the shell degeneracy

\begin{equation}
    N_k \leq \Omega_\mu,
\end{equation}
with generic choices of coefficients $c_{\mathbf{k}}^{(j)}$ including all available momenta. The maximum shell degeneracy below some arithmetic norm $M$ grows subpolynomially \cite{https://doi.org/10.1112/blms.12804}

\begin{equation}
    \Omega_\mu \leq \Omega_{\max}(M)=M^{o(1)}=L^{o(1)},
    \qquad m_\mu \leq M,
\end{equation}
since $M \sim L^2$. The number of momenta $N_k$ carried by a state in the defect-decoupled sector is therefore given by

\begin{equation}
    N_k \leq L^{o(1)}.
\end{equation}

Around a skyrmion, the defect-coupled sector supports hybridization of magnon wavefunctions with magnetically active internal deformations, and is therefore of interest. The perturbative mixing parameter between defect-coupled states in different shells is

\begin{equation}
    \eta_{\mu\mu'}
    \sim
    \frac{
    \left|\langle B_\mu|\hat V|B_{\mu'}\rangle\right|
    }{
    |\omega_\mu-\omega_{\mu'}|
    }
    =
    \frac{|U|}{\lambda}
    \frac{\sqrt{\Omega_\mu\Omega_{\mu'}}}
    {|m_\mu-m_{\mu'}|}.
\end{equation}
Strong inter-shell hybridization is expected when $\eta_{\mu\mu'}\gtrsim 1$, corresponding to

\begin{equation}
    |m_\mu-m_{\mu'}| \lesssim \frac{|U|}{\lambda} \sqrt{\Omega_\mu\Omega_{\mu'}} \leq L^{o(1)}.
\end{equation}
To show that mesoscopic multifractality is permitted, it is sufficient to retain only these strongly hybridized shells. Since this window contains at most $L^{o(1)}$ shells, each with at most $L^{o(1)}$ momenta, the number of retained momenta satisfies

\begin{equation}
    N_k \leq L^{o(1)} L^{o(1)} = L^{o(1)}.
\end{equation}

Denoting the set of these momenta as $\mathcal{S}$, the wavefunction can be written as

\begin{equation}
    \psi(\mathbf{r})
    =
    \frac{1}{L}
    \sum_{\mathbf{k}\in \mathcal{S}}
    c_{\mathbf{k}}
    e^{i\mathbf{k}\cdot\mathbf{r}},
    \qquad
    \sum_{\mathbf{k}\in \mathcal{S}}
    |c_{\mathbf{k}}|^2=1,
    \qquad
    |\mathcal{S}|=N_k.
    \end{equation}
The Cauchy-Schwarz inequality provides an upper bound to the local probability and its logarithm

\begin{equation}\label{CS}
    |\psi_i|^2 \leq \frac{N_k}{L^2}, \quad -\log{|\psi_i|^2} \geq -\log{\frac{N_k}{L^2}}.
\end{equation}
We now introduce the Rényi entropy
\begin{equation}
    S_q=\frac{1}{1-q}\log\sum_i |\psi_i|^{2q},
\end{equation}
which reduces to the Shannon entropy
$S_1=-\sum_i |\psi_i|^2\log|\psi_i|^2$
in the limit $q\rightarrow1$, and is related to the generalized dimensions by
\begin{equation}
    D_q=\frac{S_q}{\log L}.
\end{equation}
Using Eq. \eqref{CS} and normalization of the wavefunction,
\begin{equation}
    \sum_i |\psi_i|^{2q}
    \begin{cases}
        \leq \left(N_k/L^2\right)^{q-1}, & q>1,\\[4pt]
        \geq \left(N_k/L^2\right)^{q-1}, & 0<q<1.
    \end{cases}
\end{equation}
Since the prefactor $1/(1-q)$ changes sign at $q=1$, both cases imply
\begin{equation}
    S_q \geq 2\log L-\log N_k .
\end{equation}
The Rényi entropy of a normalized distribution over $L^2$ sites is
bounded above by $2\log L$, and therefore
\begin{equation}
    2-\frac{\log N_k}{\log L}
    \leq D_q \leq 2,
    \qquad q>0,
\end{equation}
where the $q=1$ result follows from the continuous limit. The approach to the extended limit is permitted to be exceptionally
slow, which is apparent after substituting the sharper bound
$\log\Omega_{\max}(M)=O(\log M/\log\log M)$ \cite{https://doi.org/10.1112/S0025579319000214}. This implies
\begin{equation}
    \log N_k=O\left(\frac{\log L}{\log\log L}\right),
\end{equation}
and therefore
\begin{equation}
    2-O\left(\frac{1}{\log\log L}\right)
    \leq D_q\leq2,
    \qquad q>0.
\end{equation}

\paragraph{Continuum limit.} We next determine the positive-order multifractal dimensions of the continuum magnon--skyrmion scattering states. The continuum modes satisfy
the bosonic spin-wave eigenvalue problem $\mathcal{H}\widetilde{\psi} =\varepsilon\tau_z\widetilde{\psi}$, where $\widetilde{\psi}=(u,v)^T$ contains the particle and hole components, as discussed in \cite{PhysRevB.90.094423}. Rotational symmetry allows the partial-wave decomposition $\widetilde{\psi}_m(r,\chi)=e^{im\chi}\widetilde{\eta}_m(r)$. For an
above-gap state, $\varepsilon=\varepsilon_{\rm gap}+\hbar^2k^2/(2M_{\rm mag})$ with $k>0$. The skyrmion scattering potential is exponentially localized on the scale $R_{\rm sk}$. The regular partial-wave solutions behave near the origin as $u_m\sim r^{|m+1|}$ and $v_m\sim r^{|m-1|}$ and therefore have finite local intensity. At distances $r\gg R_{\rm sk}$, the hole component is exponentially suppressed, and the particle component is a linear combination of Bessel functions, $u_m\propto \cos\delta_m J_{m-1}(kr)-\sin\delta_m Y_{m-1}(kr)$. Equivalently, a plane-wave scattering state in the laboratory frame has the asymptotic form
\begin{equation}
\widetilde{\psi}_{\rm LAB}(\mathbf r)=
\begin{pmatrix}
1 \\ 0
\end{pmatrix}
\left[e^{i\mathbf{k}\cdot\mathbf r}+f(\chi)\frac{e^{ikr}}{\sqrt{r}}\right]
+O(r^{-3/2})+
\begin{pmatrix}
0 \\ O(e^{-r/R_{\rm sk}})
\end{pmatrix}.
\label{eq:continuum_scattering_state}
\end{equation}
The phase shifts $\delta_m$, including their asymmetry under reversal of the angular momentum, determine the angular scattering amplitude $f(\chi)$ and the differential cross section \cite{PhysRevB.90.094423}. They modify the spatial interference pattern but do not generate singular or lower-dimensional support.

Continuum scattering states are delta-normalized on the infinite plane. To apply the multifractal diagnostics, we restrict the state to a finite physical observation region $\Omega$ and normalize the same positive intensity measure used for the lattice Bogoliubov modes. Defining $w(\mathbf r)=|u(\mathbf r)|^2+|v(\mathbf r)|^2$, we introduce
\begin{equation}
\varrho(\mathbf r)=\frac{w(\mathbf r)}
{\displaystyle\int_{\Omega}d^2r\,w(\mathbf r)},
\qquad
\int_{\Omega}d^2r\,\varrho(\mathbf r)=1.
\label{eq:continuum_probability}
\end{equation}
The regular short-distance behavior and the smooth localized scattering potential imply that $\varrho(\mathbf r)$ is a bounded, continuous function on every finite observation region.

We partition $\Omega$ into boxes $B_n(\ell)$ of side length $\ell$ and
define $P_n(\ell)=\int_{B_n(\ell)}d^2r\,\varrho(\mathbf r)$. For every
bulk box, continuity of $\varrho$ implies
$P_n(\ell)=\varrho(\mathbf r_n)\ell^2+o(\ell^2)$ for some
$\mathbf r_n\in B_n(\ell)$. Boundary boxes give only subleading
corrections. For $q>0$, the partition sum therefore becomes
\begin{equation}
    Z(q,\ell)
    \sum_n P_n(\ell)^q=
    \ell^{2q}
    \sum_n
    \varrho(\mathbf r_n)^q
    \left[1+o(1)\right]=
    \ell^{2(q-1)}
    \left[
        \ell^2
        \sum_n\varrho(\mathbf r_n)^q
    \right]
    \left[1+o(1)\right].
\label{eq:continuum_partition_sum_derivation}
\end{equation}
The expression in square brackets is a Riemann sum and approaches
$\int_{\Omega}d^2r\,\varrho(\mathbf r)^q$, which is finite for every
$q>0$. Introducing the dimensionless box size
$\lambda=\ell/(a_0 L)$, the partition sum has the asymptotic form
$Z(q,\lambda)=C_q\lambda^{2(q-1)}[1+o(1)]$, where $C_q$ is independent of
$\lambda$. Consequently,
\begin{equation}
    \tau(q)=2(q-1),
    \qquad
    D_q=\frac{\tau(q)}{q-1}=2,
    \qquad
    q>0,\quad q\neq1.
\label{eq:continuum_dimensions}
\end{equation}
The information dimension follows from the continuous $q\rightarrow1$
limit of the generalized dimensions already defined in the main text,
$D_1=\lim_{q\rightarrow1}D_q=\tau'(1)$. Hence $D_1=2$, while
Eq.~\eqref{eq:continuum_dimensions} gives in particular $D_2=2$.

For the same positive-$q$ sector, the singularity strength and its
Legendre transform are
\begin{equation}
    \alpha(q)=\frac{d\tau(q)}{dq}=2,
    \qquad
    f(\alpha(q))=q\alpha(q)-\tau(q)=2,
    \qquad q>0.
\label{eq:continuum_singularity_spectrum}
\end{equation}
Thus, the positive-$q$ branch of the continuum singularity spectrum collapses to the single point $(\alpha,f)=(2,2)$. The continuum magnon--skyrmion scattering states are therefore extended states according to all positive-order generalized dimensions, including the information and correlation dimensions $\mathcal{D}_1$ and $\mathcal{D}_2$.

\newpage

\subsection{E. Numerical Scaling of Multifractal Diagnostics}

Figure~\ref{D1_dependence} shows the system-size dependence of the multifractal diagnostics for chiral magnon modes around target frequency $\omega = 1.4$ for fixed Hamiltonian parameters and skyrmion size. Over the accessible range, the information dimension $\mathcal{D}_1$ and correlation dimension $\mathcal{D}_2$ both remain below $D_0$, mass exponents $\tau(q)$ retain their nonlinearity, and spectral width $\Delta \alpha$ stays separated from its extended-limit value of $\Delta \alpha=0$. Both dimensions drift monotonically toward $D_0$ with increasing $L$, as expected from Appendix D, while remaining below it by $0.05–0.15$ over the full range studied. 

\begin{figure}[H]
\centering
\includegraphics[width=0.85\linewidth]{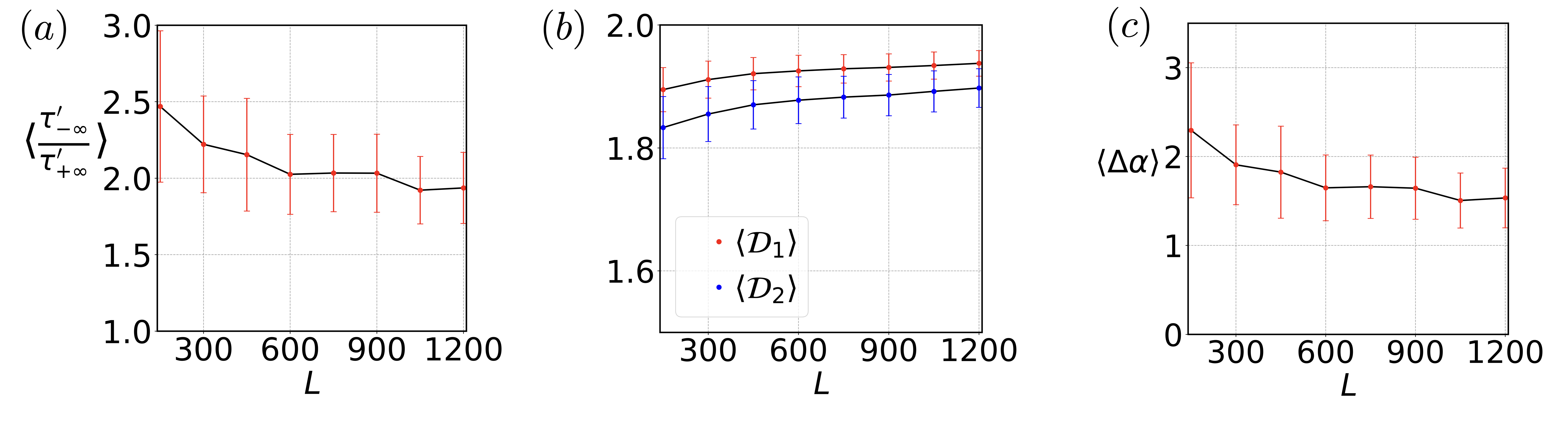}
\caption{\textbf{System Size Scaling.} Multifractal signals averaged over $100$ chiral modes around target energy $\omega = 1.4$ persist over mesoscopic sizes. (a) Nonlinearity $\tau'_{-\infty}/\tau'_{+\infty}$ of $\tau(q)$ measured as the ratio of slopes in the two limits $q \rightarrow -\infty,\,+\infty$ remains far from the extended limit $\tau'_{-\infty}/\tau'_{+\infty} = 1$. (b) Information and correlation dimensions $\mathcal{D}_1$ and $\mathcal{D}_2$ below the extended limit of $\mathcal{D}_1 = \mathcal{D}_2 = D_0$. (c) Spectral width $\Delta \alpha$ above the monofractal limit $\Delta \alpha = 0$. Error bars display standard deviations.}
\label{D1_dependence}
\end{figure}

We also test the dependence of multifractal signatures on skyrmion size $R_{\rm sk}$, obtained by varying Hamiltonian parameters $D$ and $h$ at fixed reduced field $\beta = hJ/D^2=1.2$. The weak dependence shows that the multifractal response persists around substantially larger skyrmions than the lattice-scale skyrmions used throughout the main calculations. 

\begin{figure}[H]
\centering
\includegraphics[width=0.85\linewidth]{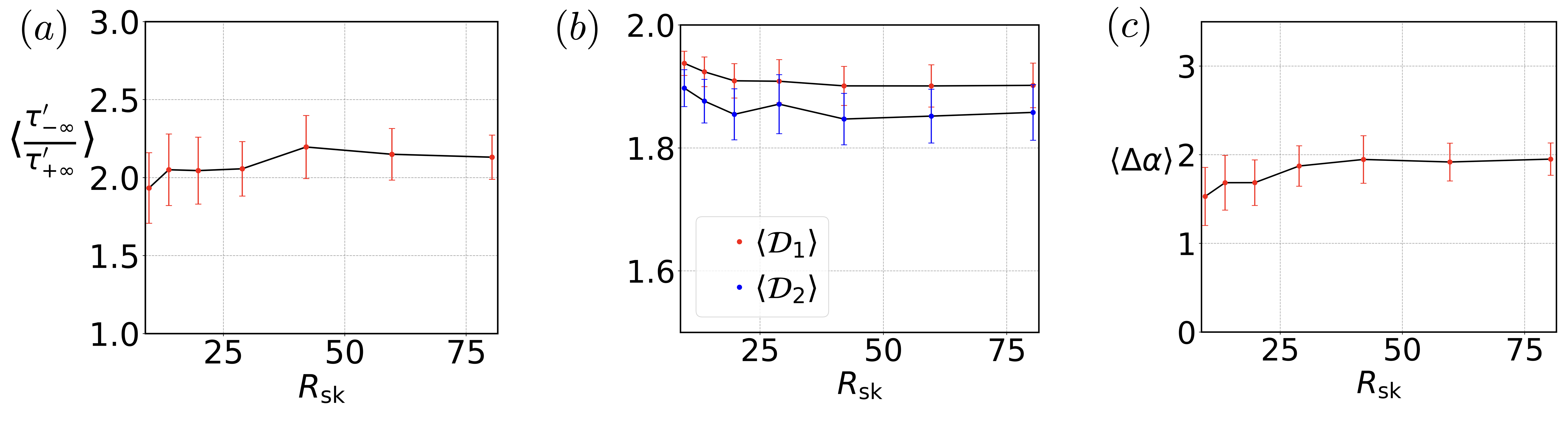}
\caption{\textbf{Skyrmion Size Scaling.} Multifractal signals averaged over $100$ chiral modes around target energy $\omega = \omega_{\rm gap} + 0.2 D^2/J$. The multifractal diagnostics depend only weakly on skyrmion size and remain separated from their extended-state limits. Skyrmion size is varied by changing Hamiltonian parameters $h$ and $D$ with fixed $\beta = hJ/D^2=1.2$. (a) Nonlinearity $\tau'_{-\infty}/\tau'_{+\infty}$ of $\tau(q)$ measured as the ratio of slopes in the two limits $q \rightarrow -\infty,\,+\infty$ remains far from the extended limit $\tau'_{-\infty}/\tau'_{+\infty} = 1$. (b) Information and correlation dimensions $\mathcal{D}_1$ and $\mathcal{D}_2$ below the extended limit of $\mathcal{D}_1 = \mathcal{D}_2 = D_0$. (c) Spectral width $\Delta \alpha$ above the monofractal limit $\Delta \alpha = 0$. Error bars display standard deviations.}
\label{L_SK_dependence}
\end{figure}

\subsection{F. Generic Multifractality around Local Defects}

The defect-scattering picture suggests that multifractal structure can appear generically across modes. We verify this numerically by computing eigenstates around the $S=0$ defect considered in Fig.~\ref{skyrmion_dynamics} near several energy targets. Fig.~\ref{Scatterplot_Defect} shows the resulting distributions of $\Delta \alpha$ and $\mathcal{D}_1$, confirming that multifractal signatures occur broadly across the sampled spectrum.

\begin{figure}[H]
\centering
\includegraphics[width=0.3\linewidth]{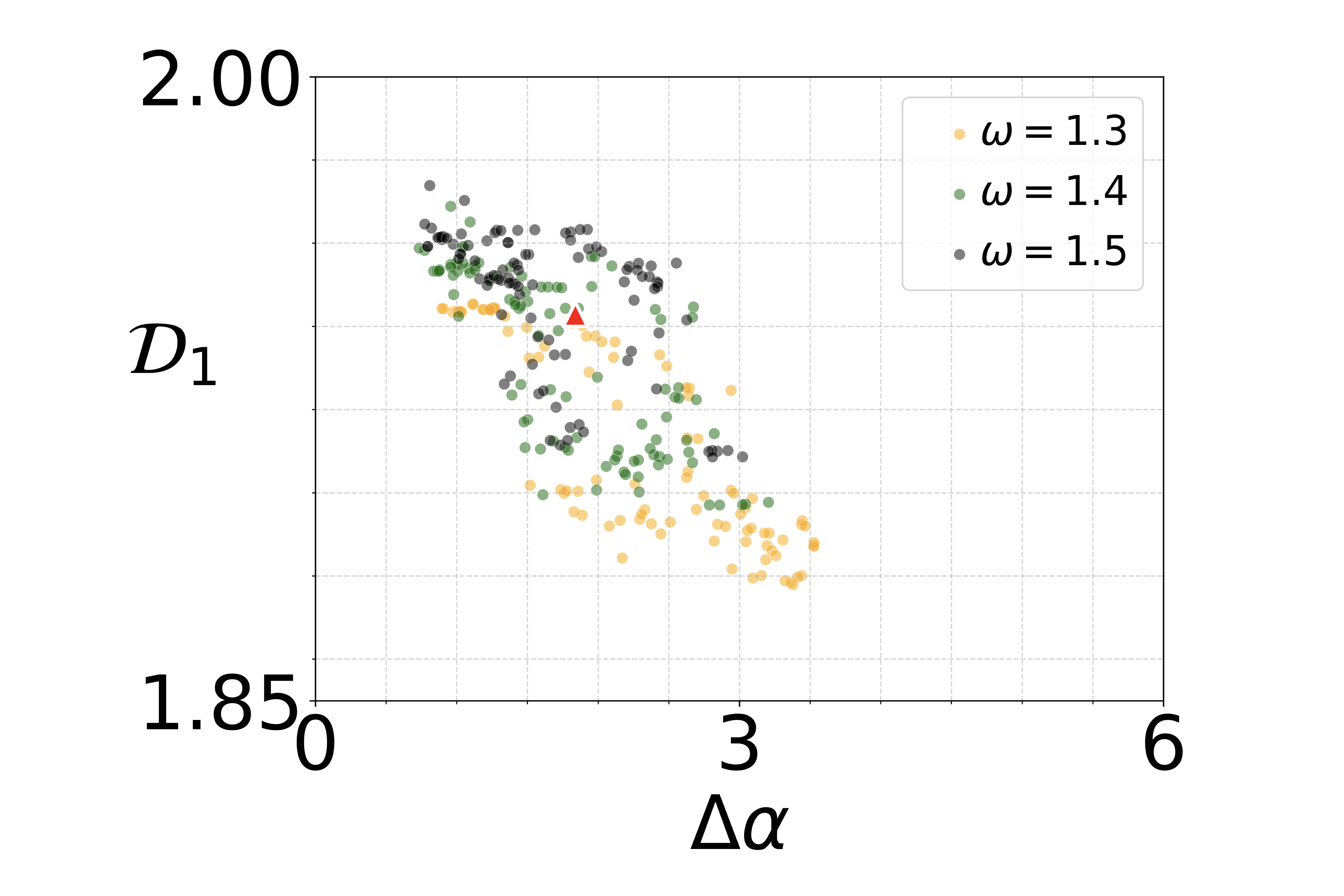}
\caption{\textbf{Generic Multifractality around Defects.} Scatterplots of $\Delta \alpha$ and $\mathcal{D}_1$ for modes found nearby different energy targets around a single $S=0$ spin defect, illustrating generic multifractality. The red triangle corresponds to the mode displayed in Fig. \ref{skyrmion_dynamics}.}
\label{Scatterplot_Defect}
\end{figure}

\subsection{G. Multifractality around Multiple Defects}

The eigenstates of magnons scattered by a single defect are characterized by coherent interference patterns centered on the scatterer, with the number of participating momenta depending on the defect energy and system size. In systems with multiple defects, additional scattering can hybridize a greater number of momentum states while progressively scrambling this real-space structure. We numerically test the dependence of multifractal signatures on defect number in the skyrmion system. Figure~\ref{Multiple_Defects_Appendix} shows the dependence of the information and correlation dimensions for different numbers of randomly-placed skyrmions, together with representative real-space wavefunctions. Multifractality is enhanced by skyrmion number, but the effect saturates, suggesting competition between enhanced shell hybridization and the suppression of multifractality by phase scrambling.

\begin{figure}[H]
\centering
\includegraphics[width=0.9\linewidth]{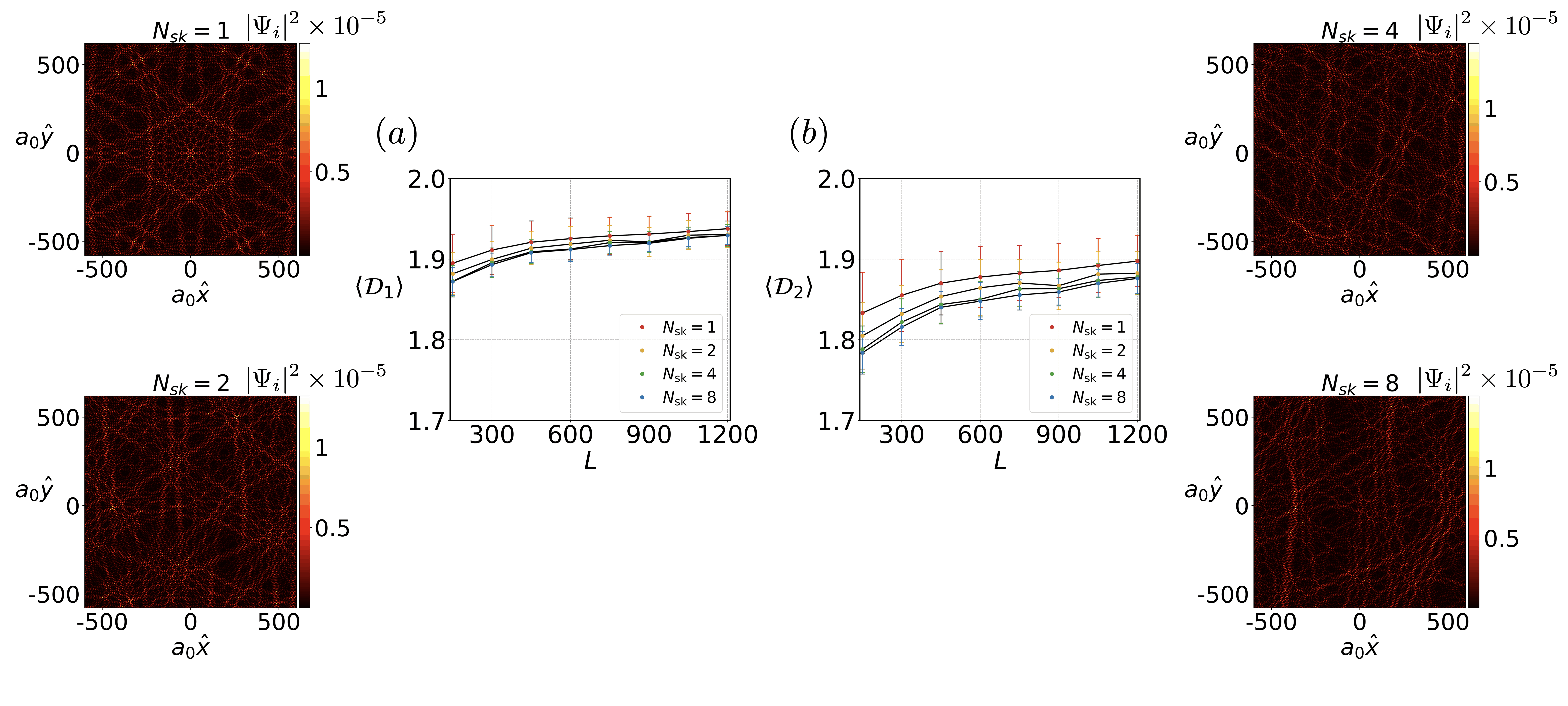}
\caption{\textbf{Modes around Multiple Defects.} (a) Information dimensions $\mathcal{D}_1$ and (b) correlation dimensions $\mathcal{D}_2$ averaged over $100$ chiral magnon modes for varying skyrmion numbers. Each curve corresponds to a fixed number $N_{\rm sk}$ of randomly positioned skyrmions as the linear system size $L$ is varied. Error bars display standard deviations. Surrounding images illustrate real-space structure of representative wavefunctions.}
\label{Multiple_Defects_Appendix}
\end{figure}

To examine the phase structure of multiple-defect scattering, we consider $N_d$ point defects at locations $\mathbf{r}_a$, for which the momentum-space matrix elements are

\begin{equation}
    V_{\mathbf k,\mathbf k'}
    = \frac{U}{L^2}
    \sum_{a=1}^{N_d}
    e^{i(\mathbf k'-\mathbf k)\cdot\mathbf r_a}.
\end{equation}
The momentum-dependent phases introduced by additional defects modify coherent interference, motivating a phase-scrambling test of the single-shell wavefunction in Eq. \eqref{coherent_wavefunction}. Introducing a Gaussian random phase shift $\phi_{\mathbf k}\sim\mathcal{N}(0,\sigma^2)$, the scrambled test wavefunction reads

\begin{equation}
    \psi_i=\frac{1}{\sqrt{N\Omega_\mu}}
    \sum_{\mathbf{k}\in S_\mu}
    e^{i\mathbf{k}\cdot(\mathbf{r}_i-\mathbf{r}_0)+i\phi_{\mathbf{k}}}.
\end{equation}
Figure \ref{Phase_Randomization} shows that increasing $\sigma$ weakens but does not eliminate multifractal signatures. 

\begin{figure}[H]
\centering
\includegraphics[width=0.7\linewidth]{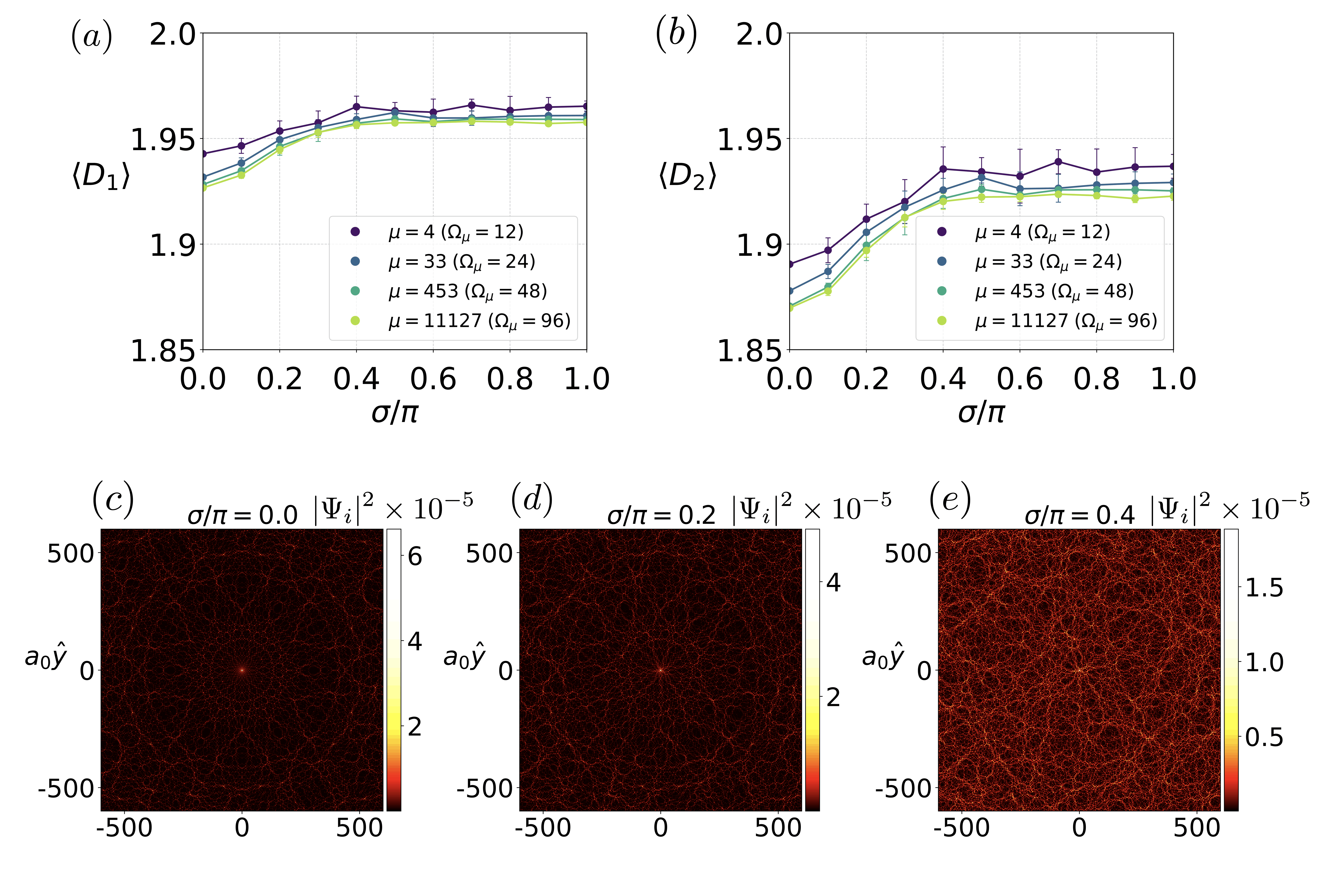}
\caption{\textbf{Multifractality in Phase-Scrambled Wavefunctions.} (a) Information dimensions $\mathcal{D}_1$ and (b) correlation dimensions $\langle \mathcal{D}_2 \rangle$ (b) as functions of the random phase-disorder strength $\sigma$, averaged over $10$ reconstructed superpositions of shell states from selected shells with $\Omega_\mu=12,24,48,96$. The wavefunctions are constructed as $\psi_i=(1/\sqrt{N\Omega_\mu})\sum_{\mathbf k\in S_\mu}
e^{i\mathbf k\cdot(\mathbf r_i-\mathbf r_0)+i\phi_{\mathbf k}}$, where $\phi_{\mathbf k}\sim\mathcal{N}(0,\sigma^2)$ are independent Gaussian random variables with zero mean and variance $\sigma^2$. Error bars display the standard deviation. Phase disorder suppresses multifractality, but signatures saturate below the extended limit. (c)-(e) Representative wavefunctions at varying levels of phase-disorder, illustrating the transition from coherent to scrambled multifractal spatial structure.}
\label{Phase_Randomization}
\end{figure}

\subsection{H. Finite-Resolution Observability}

Figure~\ref{Observability_Appendix} repeats the finite-resolution analysis reported in Fig~\ref{FractDimens}(a) for smaller and larger frequency-bin widths. Across the resolutions shown, windows carrying finite dipole weight continue to exhibit dipole-weighted information dimensions lower than $D_0$. The multifractal response therefore remains observable across varying frequency resolution.

\begin{figure}[H]
    \centering
    \includegraphics[width=0.7\linewidth]{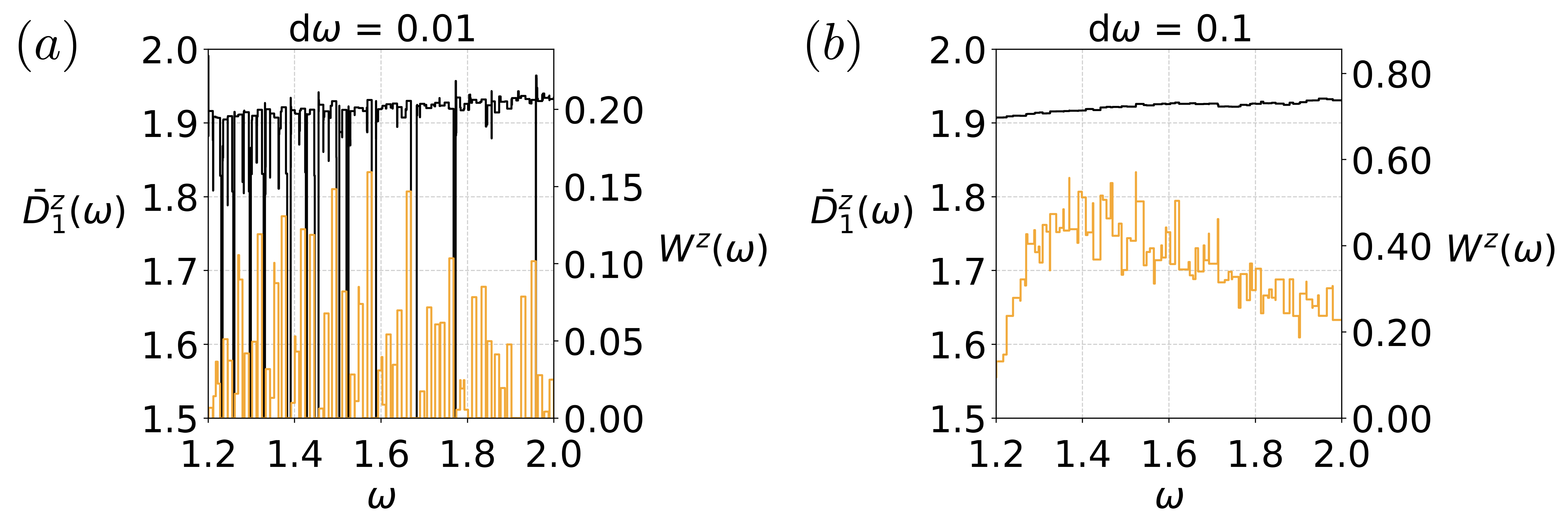}
    \caption{\textbf{Observability for Different $\boldsymbol{\delta}\boldsymbol{\omega}$}. Longitudinal dipole spectral weight $W^{z}(\omega)$ (orange, right axis) and dipole-weighted information dimension $\overline{D}^{\,z}_1(\omega)$ (black, left axis) for varying window sizes, illustrating robust observability of multifractal above-gap modes.}
    \label{Observability_Appendix}
\end{figure}

\subsection{I. Spin-Current Matrix Elements for Magnons}

\begin{figure}[H]
    \centering
    \includegraphics[width=0.7\linewidth]{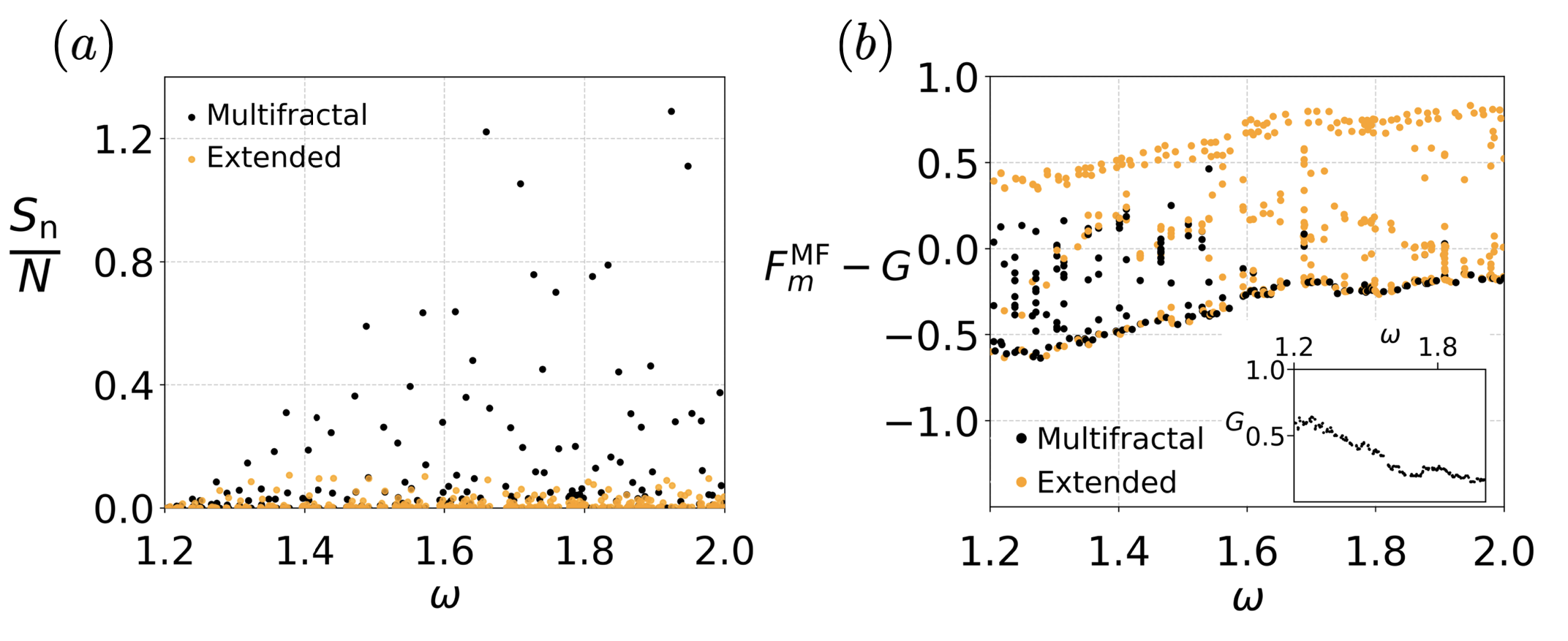}
    \caption{(a) Normalized spin-current weight toward multifractal-like ($\mathcal{D}_1 < 1.9$) states is typically about an order of magnitude larger than toward extended-like ($\mathcal{D}_1 \geq 1.9$) states. (b) The multifractal current fraction $F^{\rm MF}_m$ relative to phase-space fraction $G$ (inset) forms two branches characterized by preferential coupling either to multifractal-like or extended-like modes. Points are colored by source-mode category, with multifractal-like modes in black and extended-like modes in orange, revealing a statistical tendency toward cross-sector connectivity.}
    \label{fig:SpinCurrent}
\end{figure}

We summarize here the derivation of the quantities reported in Fig. \ref{fig:SpinCurrent}, which starts from the Heisenberg equation of motion for the local $z$-spin density:
\begin{equation}
\dot S_i^z=\frac{i}{\hbar}[H,S_i^z].
\end{equation}

\noindent
The Hamiltonian can be decomposed into on-site and bond terms:

\begin{equation}
H = \sum_i H_i^{\rm on} + \sum_{ij} H_{ij}.
\end{equation}

\noindent
We define the bond spin current spin-$z$ flowing from site $i$ to $j$ by
\begin{equation}
J_{i\to j}^z \equiv -\frac{i}{\hbar}\,[H_{ij},S_i^z].
\label{eq:defJ}
\end{equation}

\noindent
Then, the equation of motion can be written as the lattice continuity equation
\begin{equation}
\dot S_i^z + \sum_{j \in \mathcal{N}_i} J_{i\to j}^z = \tau_i^z,
\label{eq:continuity}
\end{equation}

\noindent
where $\mathcal{N}_i$ denotes the set of sites coupled to $i$ by bond terms and
\begin{equation}
\tau_i^z = \frac{i}{\hbar}\,[H_i^{\rm on},S_i^z]
\end{equation}

\noindent
is a local torque/source term which vanishes if the terms commute with $S_i^z$. For Heisenberg bond terms $H_{\langle ij \rangle}^{\rm ex}=J\,\textbf{S}_i\cdot \textbf{S}_j$ of eq. \eqref{eq:HamChiral}, we compute:

\begin{align}
[H_{ij}^{\rm ex},S_i^z]
&= J_{ij}\Big([S_i^x,S_i^z]S_j^x + [S_i^y,S_i^z]S_j^y\Big) \nonumber\\
&= J_{ij}\Big((-i\hbar S_i^y)S_j^x + (i\hbar S_i^x)S_j^y\Big)
= i\hbar J_{ij}\,(S_i^x S_j^y - S_i^y S_j^x).
\end{align}

\noindent
Using the definition \eqref{eq:defJ}, the bond current is therefore
\begin{equation}
J_{i\to j}^{z,{\rm ex}}=
J_{ij}\,(S_i^x S_j^y - S_i^y S_j^x)=
J_{ij}\,(\mathbf{S}_i \times \mathbf{S}_j)_z.
\label{eq:Jex}
\end{equation}

\noindent
DMI bond terms $H_{ij}^{\rm DM} = \mathbf{D}_{ij}\cdot(\mathbf{S}_i\times \mathbf{S}_j)$ yield:

\begin{align}
[H_{ij}^{\rm DM},S_i^z]
&=
D_{ij}^x [S_i^y,S_i^z]S_j^z
- D_{ij}^y [S_i^x,S_i^z]S_j^z
+ D_{ij}^z\Big([S_i^x,S_i^z]S_j^y - [S_i^y,S_i^z]S_j^x\Big)\nonumber\\
&=
i\hbar\Big(
D_{ij}^x S_i^x S_j^z
+ D_{ij}^y S_i^y S_j^z
- D_{ij}^z (S_i^x S_j^x + S_i^y S_j^y)
\Big).
\end{align}

\noindent
The total bond current for the chiral model \eqref{eq:HamChiral} is then
\begin{equation}
J_{i\to j}^{z}
=
J_{i\to j}^{z,{\rm ex}} + J_{i\to j}^{z,{\rm DM}}
=
J\,(S_i^x S_j^y - S_i^y S_j^x)
+
D_{ij}^x S_i^x S_j^z
+ D_{ij}^y S_i^y S_j^z
- D_{ij}^z (S_i^x S_j^x + S_i^y S_j^y).
\label{eq:Jchiral}
\end{equation}
Since $S^z$ is not conserved, 
\eqref{eq:continuity} holds with a nonzero $\tau_i^z$.
Given a lattice direction $\hat e_\mu$,
a standard definition of the total spin current operator is
\begin{equation}
\mathcal{J}_\mu^z
\equiv \sum_{\langle ij\rangle} \eta_{ij}\, J_{i\to j}^z,
\qquad
\eta_{ij}\equiv (\bm r_j-\bm r_i)\cdot \hat{\bm e}_{\mu},
\end{equation}

\noindent
For our calculations, we set $\mu = x$

To represent $\mathcal{J}_x^z$ in the magnon Hilbert space, we expand 
$\bm S_i=\bm S_{i,0}+\delta\bm S_i$ in the local rotating frame 
$\{\bm v_i^{1},\bm v_i^{2},\bm v_i^{3}\}$ with $\bm v_i^{3}\parallel \bm S_{i,0}$,
and express $\delta\bm S_i$ in terms of HP bosons $a_i,a_i^\dagger$:
\begin{equation}
S_i^{3}=S-a_i^\dagger a_i,\qquad
S_i^{1}=\sqrt{\frac{S}{2}}\,(a_i+a_i^\dagger),\qquad
S_i^{2}=-i\sqrt{\frac{S}{2}}\,(a_i-a_i^\dagger).\label{eq:HP_operators}
\end{equation}
It is convenient to define the local-frame coefficients (depending only on the texture)
\begin{equation}
u_{i,\alpha}\equiv v_{i}^{1,\alpha}-i v_{i}^{2,\alpha},
\qquad
\bar u_{i,\alpha}\equiv v_{i}^{1,\alpha}+i v_{i}^{2,\alpha}=u_{i,\alpha}^\ast,
\qquad
c_{i,\alpha}\equiv v_{i}^{3,\alpha},\label{HP_frame}
\end{equation}
so that
\begin{equation}
S_i^\alpha
=
c_{i,\alpha}\,(S-a_i^\dagger a_i)
+\sqrt{\frac{S}{2}}\Big(u_{i,\alpha}\,a_i+\bar u_{i,\alpha}\,a_i^\dagger\Big)
\;+\;O(S^{-1/2}).
\label{eq:Salpha_HP}
\end{equation}
Keeping terms up to quadratic order in $(a,a^\dagger)$ (linear spin waves) yields a quadratic bosonic operator
\begin{equation}
\mathcal{J}_x^z \;=\; \mathrm{const} 
+ \sum_{i,j} \Big[ a_i^\dagger\,J^{(11)}_{ij}\,a_j 
+ \frac12\big(a_i^\dagger\, J^{(12)}_{ij}\,a_j^\dagger + \mathrm{h.c.}\big)\Big]
\label{eq:J_quadratic_real}
\end{equation}

\noindent
where: 
\begin{align}
J^{(11)}_{ij}\big|_{(i\to j)}
&=
\eta_{ij}\,\frac{S}{2}\Big[
J\big(\bar u_{i,x}u_{j,y}-\bar u_{i,y}u_{j,x}\big)
+ D_{ij}^x\,\bar u_{i,x}u_{j,z}
+ D_{ij}^y\,\bar u_{i,y}u_{j,z}
- D_{ij}^z\big(\bar u_{i,x}u_{j,x}+\bar u_{i,y}u_{j,y}\big)
\Big],
\label{eq:J11_chiral}
\\[2pt]
J^{(12)}_{ij}\big|_{(i\to j)}
&=
\eta_{ij}\,\frac{S}{2}\Big[
J\big(\bar u_{i,x}\bar u_{j,y}-\bar u_{i,y}\bar u_{j,x}\big)
+ D_{ij}^x\,\bar u_{i,x}\bar u_{j,z}
+ D_{ij}^y\,\bar u_{i,y}\bar u_{j,z}
- D_{ij}^z\big(\bar u_{i,x}\bar u_{j,x}+\bar u_{i,y}\bar u_{j,y}\big)
\Big].
\label{eq:J12_chiral}
\end{align}

\noindent
The current $J_{nm}$ and intensity $I_{nm}$ between two magnon eigenstates $\ket{n},\, \ket{m}$ is given by:

\begin{equation}
J_{nm} \equiv \langle n|\mathcal{J}_x^z|m\rangle,
\qquad
I_{nm}\equiv |J_{nm}|^2.
\label{eq:Idef}
\end{equation}

\noindent
We account for spectroscopic resolution by introducing an energy neighborhood around each mode $m$:

\begin{equation}
\mathcal{W}_m(\delta)\equiv\{\,n:\ |\omega_n-\omega_m|<\delta\,\},
\label{eq:window}
\end{equation}

\noindent
where $\delta$ is a frequency resolution scale. We can now define a window-summed current coupling 
\begin{equation}
S_m(\delta)\equiv \sum_{n\in\mathcal{W}_m(\delta)} I_{nm}.
\label{eq:Sm}
\end{equation}

\noindent
Fixing a multifractality threshold $D_{1,n}<D_0-\varepsilon$, with $D_0=2$ and $\epsilon=0.1$, the current coupling can be split into multifractal-like and extended-like components:

\begin{equation}
S_m^{\rm MF}(\delta)=\sum_{\substack{n\in\mathcal{W}_m(\delta) \\ n \in \rm MF}} I_{nm},
\qquad
S_m^{\rm E}(\delta)=\sum_{\substack{n\in\mathcal{W}_m(\delta) \\ n \in \rm E}} I_{nm},
\end{equation}

\noindent
which we normalized by the available phase space (number of modes) for each type $N^{\text{MF}}$, $N^{\text{E}}$. The fraction of the spin current weight directed toward multifractal states is given by:

\begin{equation}
F_m^{\rm MF}(\delta)\equiv
\frac{S_m^{\rm MF}(\delta)}{S_m^{\rm MF}(\delta)+S_m^{\rm E}(\delta)},
\label{eq:FMF}
\end{equation}

\noindent
which can be compared to the fraction of phase space occupied by multifractal modes to identify a bias:

\begin{equation}
    G = \frac{N^{\text{MF}}}{N^{\text{E}} + N^{\text{MF}}}.
\end{equation}

We have observed that multifractality leaves signatures in transport-like observables. Dynamical connectivity between single-magnon states is encoded in spin-current matrix elements $J_{nm}\equiv\langle n|\hat J^z_\mu|m\rangle$ and intensities $I_{nm}=|J_{nm}|^2$. At finite resolution $\delta \omega$, we sum $I_{nm}$ over states within $|\omega_n - \omega_m|<\delta \omega$ to obtain the window-summed coupling $S_m(\delta \omega)$, which is separated into contributions $S_m^{\rm MF}(\delta \omega)$ and $S_m^{\rm E}(\delta \omega)$ from multifractal-like ($\mathcal{D}_1 < D_0 - \epsilon$) and relatively less multifractal extended-like ($\mathcal{D}_1 \geq D_0 - \epsilon$) sectors. Normalizing this coupling by the number of states $N^{\rm MF, E}$ in each sector within the window removes the trivial phase-space imbalance, allowing a direct comparison of coupling per available state. Fig.~\ref{fig:SpinCurrent}(a) reports $S_m^{\rm E}/N^{\rm E}$ and $S_m^{\rm MF}/N^{\rm MF}$ for $\delta \omega = 0.05$ and $\epsilon = 0.1$ calculated from the full magnon spectrum of a $120 \times 120$ lattice, showing that coupling toward multifractal-like states is typically around one order of magnitude higher than toward extended-like states. 

We further characterize magnon connectivity by the fraction of current weight directed toward the multifractal branch $F^{\rm MF}_m$, and $G$, the phase space fraction occupied by multifractal modes in the window. The bias $F^{\rm MF}_m - G$ separates into two branches associated with preferential coupling toward either multifractal-like or extended-like sectors. Resolving by the character of the source mode reveals a statistical tendency of extended-like modes to bias toward multifractal-like couplings and for multifractal-like modes to bias toward extended-like couplings, indicating cross-connectivity between the two sectors.

\subsection{J. Examples of Dipole-Active Modes}

Figure~\ref{Observable_Gallery} shows the multifractal modes with the largest dipole moments obtained around chiral and frustrated skyrmions near several energy targets. The observable states exhibit diverse spatial patterns accompanied by consistent multifractal signatures. 

\begin{figure}[H]
    \centering
    \includegraphics[width=0.98\linewidth]{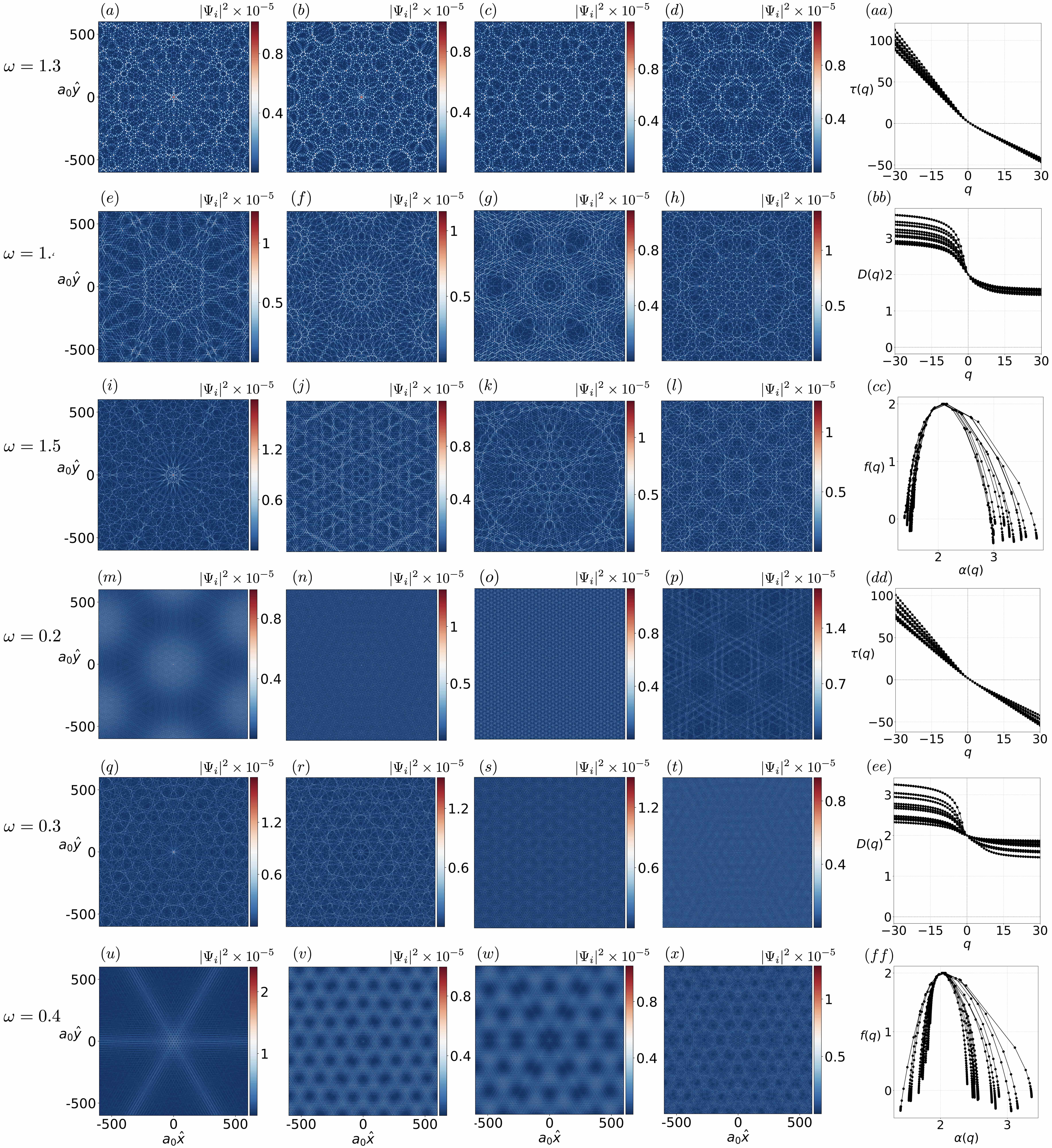}
    \caption{\textbf{Observable Multifractal Modes} Magnon wavefunctions selected by highest dipole moments obtained at various energy targets around a chiral skyrmion (a)-(l) and a frustrated skyrmion (m)-(x). (aa)-(cc) $\tau(q)$, $\mathcal{D}_q$, and $(\alpha, f(\alpha))$ for chiral magnon modes. (dd)-(ff) $\tau(q)$, $\mathcal{D}_q$, and $(\alpha, f(\alpha))$ for frustrated magnon modes.}
    \label{Observable_Gallery}
\end{figure}

\end{widetext}

\end{document}